\documentclass[twocolumn,journal]{IEEEtran}
\usepackage[T1]{fontenc}
\usepackage[utf8]{inputenc}
\usepackage{color}
\usepackage{calc}
\usepackage{textcomp}
\usepackage{amsmath}
\usepackage{graphicx}
\usepackage{multirow}

\makeatletter

\newcommand{\lyxmathsym}[1]{\ifmmode\begingroup\def\b@ld{bold}
  \text{\ifx\math@version\b@ld\bfseries\fi#1}\endgroup\else#1\fi}

\let\oldforeign@language\foreign@language
\DeclareRobustCommand{\foreign@language}[1]{%
  \lowercase{\oldforeign@language{#1}}}

\usepackage[caption=false,font=footnotesize]{subfig}

\@ifundefined{showcaptionsetup}{}{%
 \PassOptionsToPackage{caption=false}{subfig}}
\usepackage{subfig}
\makeatother

\begin{document}

\title{Methodology for accurately assessing the quality perceived by users on 360VR contents}

\author{Lara~Mu\~noz\thanks{L.~Mu\~noz, C. D\'iaz, M. Orduna, J. I. Ronda and N. Garc\'ia are with
the Grupo de Tratamiento de Im\'agenes, Information Processing and Telecommunications
Center and ETSI Telecomunicaci\'on, Universidad Polit\'ecnica de Madrid,
Madrid, Spain, e-mails: lms@gti.ssr.upm.es; cdm@gti.ssr.upm.es; moc@gti.ssr.upm.es; jir@gti.ssr.upm.es; narciso@gti.ssr.upm.es}, C\'esar~D\'iaz, Marta~Orduna, Jos\'e~Ignacio~Ronda, Pablo~P\'erez,
Ignacio~Benito\thanks{P. P\'erez and I. Benito are with Nokia Bell Labs, Madrid, Spain, e-mail: pablo.perez@nokia-bell-labs.com; ignacio.benito\_frontelo@nokia-bell-labs.com}, and Narciso~Garc\'ia}

\markboth{IEEE JSTSP Special Issue on Perception-Driven 360-Degree Video Processing}{L. Mu\~noz \MakeLowercase{\emph{et al.}}: Accurate methodology to evaluate viewport-adaptive methods for providing
360VR content}
\maketitle
\begin{abstract}
To properly evaluate the performance of 360VR-specific encoding and transmission schemes, and particularly of the solutions based on viewport adaptation, it is necessary to consider not only the bandwidth saved, but also the quality of the portion of the scene actually seen by users over time. With this motivation, we propose a robust, yet flexible methodology for accurately assessing the quality within the viewport along the visualization session. This procedure is based on a complete analysis of the geometric relations involved. Moreover, the designed methodology allows for both offline and online usage thanks to the use of different approximations. In this way, our methodology can be used regardless of the approach to properly evaluate the implemented strategy, obtaining a fairer comparison between them.
\end{abstract}


\section{Introduction}

\IEEEPARstart{D}{uring} the last few years, the interest for Virtual Reality (VR) has grown exponentially. Everyday, more and more VR-related applications appear, and the number of VR-ready devices, particularly of head-mounted displays (HMD), is quickly expanding, as they become appealing and affordable to an increasing number of users. One of the most common VR applications is the visualization via streaming of 360 videos in non-interactive environments, covering a wide range of applications such as education~\cite{freina2015literature}, medical treatments~\cite{seo2017anatomy}, and simulators~\cite{brunnstrom2018quality}.

\begin{figure}[htbp]
\begin{centering}
\textsf{\includegraphics[width=3cm]{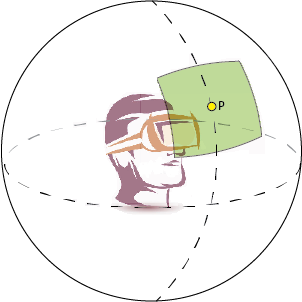}}
\par\end{centering}
\caption{\label{fig:viewport}Viewport around the point of gaze (PoG) of the
user}
\end{figure}

The transmission of this type of content is particularly challenging. The main reason is that, due to the nature of the environment and the features of the associated presentation systems, the requirements in terms of image resolution and quality to offer a really immersive experience to the user are especially demanding~\cite{allison2018perspectives,azevedo2019visual}. This results in sequences that require very high bit rates for their transmission. Thus, to provide a smooth playback and good quality service, it is necessary to incorporate coding and transmission management schemes that take advantage of the fundamental fact that only a fraction of the image can be seen by the user at a certain moment (Figure~\ref{fig:viewport}). This fraction depends on the HMD, whose design sets the field of view (FoV) and, therefore, determines the viewport, the picture area shown to the user. As the FoV is usually a right rectangular pyramid, defined by the two angles between opposing planes (dihedral angles), the projection of the viewport on the sphere is a spherical rectangle, centered on the point of gaze (PoG) of the user, that is where the user is looking at.

These coding and transmission schemes have the objective of offering high quality to the users while saving bits. It can be achieved by providing higher quality to the area that will presumably be visible to the user and lower quality to the area with a lower probability of being visible to that person. In this way, they can deliver a good quality of experience (QoE) while saving bandwidth. For this purpose, the raw image is tessellated into rectangular subimages, which are compressed and managed differently in an intelligent way. So, this procedure paves the way for the adaptation of the content presented to the viewer. In this sense, both encoding and transmission modules need to be ready to enable viewport adaptation. Regarding encoding, the technique most commonly employed to process in an independent way different areas of the image is the use of tiles, a tool included in the H.265/HEVC standard that enables the partition of the picture into independently decodable regions with some shared header information~\cite{sullivan2012overview}. Before the appearance of tiles, H.264/AVC's Flexible Macroblock Ordering (FMO)~\cite{lambert2006flexible} could be used instead to distribute heterogeneously the quality in the image. However, this tool was neither efficient nor widely implemented. Regarding transmission, HTTP/TCP-based adaptive bit rate (ABR) streaming techniques~\cite{bentaleb2018survey} are commonly used to deliver omnidirectional video, due to their adaptability. In this scheme, content is encoded at different resolutions and bit rates and divided temporally into self-contained segments of equal duration that invariably start with an Instantaneous Decoding Refresh (IDR) frame. Therefore, the segment that best suits both the system state (channel available bandwidth, terminal capabilities\dots) and the viewer's PoG at a particular moment is delivered to the client to be decoded and presented to the user. The configuration used for certain parameters, such as the number of partitions, the encoding parameter values in each image subdivision or the segment length, and the transmission scheme will influence decisively in several aspects of the system: quality provided and perceived by the user based on his/her behavior, bandwidth used, storage needs, intelligence requirements in different elements of the system, etc.

The majority of the proposed methods are evaluated considering only the bandwidth they save. For example, Ghaznavi-Youvalari et al.~\cite{ghaznavi2017viewport} compare the bandwidth saved using different strategies: SHVC, regions of interest (RoIs), etc. In the work proposed by Hosseini et al.~\cite{hosseini2017view}, the bandwidth savings of two different proposals is computed with respect to the non-viewport-adaptive version of the content. Furthermore, Zare et al.~\cite{zare2016hevc} compare the compression and bitrate performance of different grid tiling sizes. However, one key problem of viewport-adaptive approaches is not strictly related to bandwidth consumption but with the need for a system that adapts quickly to the movements of the user. That is, it is essential that the high quality viewport that corresponds to the new position of the user is presented as soon as possible. Otherwise, the user will perceive low quality areas that will decrease his/her QoE. Therefore, many strategies use short segments and small buffers. However, the drawback is that short segments imply lower coding efficiency~\cite{bedogni2017dynamic}, so a very low quality or even black areas~\cite{diaz2018viability} are necessary to be able to save bandwidth. Instead, there are some authors~\cite{sanchez2016shifted} who propose using several streams with shifted IDR frames to allow quick switching between versions prepared for different viewports and representations whereas keeping a large IDR period. However, this proposal has its own disadvantages, such as a greater complexity at the client side and a larger number of versions of the content at the server side. In either case, these changes are still not instantaneous as they require the download and playback of a new IDR, so the user could perceive low quality areas if he/she moves quickly. Furthermore, the size of the high quality areas influences significantly the QoE: the larger the areas are, the lower the probability of seeing low quality areas will be, but also the lower the bandwidth saving for an equivalent quality in the viewport.

Therefore, whichever the implemented scheme is, to measure its performance beyond the bandwidth required to transmit the content, we need a way to compute the actual quality provided to the user. In this way, we can test the design of the strategy and, when appropriate, improve it by fine-tuning the values of the parameters that characterize it. In this respect, some works measure the quality within the viewport using objective metrics, either the original version or a 360VR-aware one. In particular, Ozcinar et al.~\cite{ozcinar2017viewport} calculate both the PSNR and SSIM \cite{wang2003multiscale} metrics inside the viewport, whereas Timmerer et al.~\cite{timmerer2017adaptive} use the V-PSNR metric to determine the quality seen by the users over time. Others, such in the proposal by Xie et al.~\cite{xie2018modeling}, estimate the perceived quality considering the results of subjective tests performed previously. These experiments are focused on collecting opinion scores on different variations of the content and the resulting mean opinion score (MOS) is used to model the impact on the quality variations. Finally, there is yet another group of works that use approximations in order to determine the quality that is really perceived by the users along the session. For example, Petrangeli et al.~\cite{petrangeli2017http} use the percentage of time that the user looks to the high quality areas.

However, these proposals do not solve properly the requirements for adequately assessing  quality of omnidirectional content. Either their viewport projection is inaccurate or its implementation is not detailed enough. Hence, we present a detailed methodology, named VAQM (Viewport Adaptive Quality Method), to accurately calculate the viewport projection on the equirectangular image and, thus, to enable its use in every scheme looking for an overall quality metric value on the image seen by the user. As any standard metric (e.g. PSNR, SSIM, VMAF, MOS-related...) can be used, a complete solution for the objective quality  assessment is provided. Additionally, we also provide a simplified version of the procedure for operation under strict computing time restrictions. So, certain approximations are used for computing the quality, such as using a set of pre-calculated viewport projections.

The rest of the paper is organized as follows. First, in Section~\ref{sec:Viewport-projection}, the viewport projection is explained in detail. Then, in Section~\ref{sec:Methodology}, the procedure to obtain a figure of merit reflecting the quality perceived by the user is presented. In Section~\ref{sec:Methods}, we describe the full method and the approximated version developed to obtain the quality of the session. The description of the experiments carried out to assess the performance of the method and its results and the corresponding analysis are included in Section~\ref{sec:Results}. Finally, the paper is concluded in Section~\ref{sec:Conclusions}.

\section{\label{sec:Viewport-projection}Viewport projection}

\begin{figure}[htbp]
\begin{centering}
\textsf{\includegraphics[width=4cm]{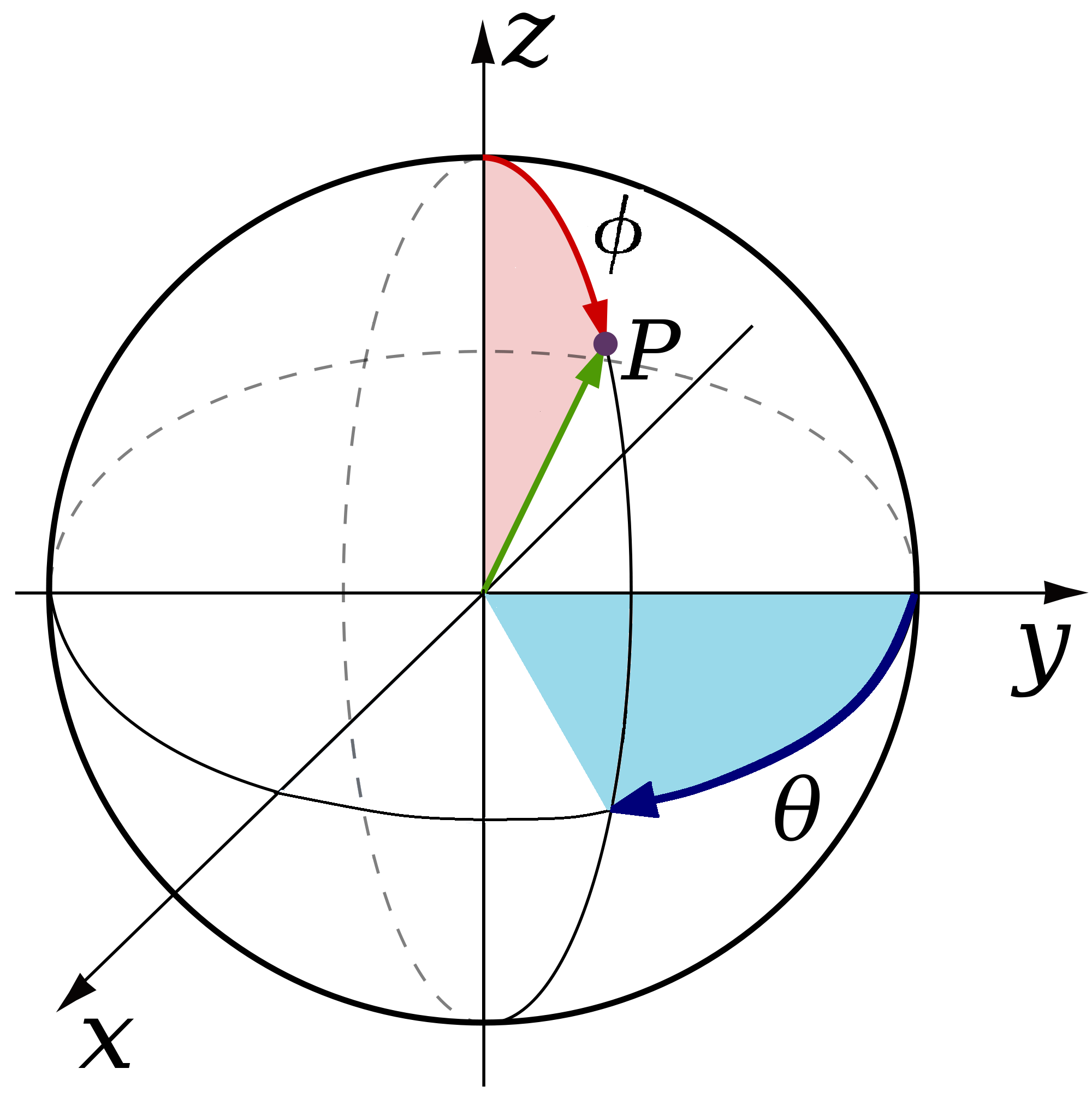}}
\par\end{centering}
\caption{\label{fig:Coordinate-System}Cartesian and spherical coordinate system on a normalized sphere}
\end{figure}

Let us consider the coordinate system used for the definition of the PoG, the FoV, and the spherical rectangle, projection of the viewport on the sphere. Figure~\ref{fig:Coordinate-System} presents the Cartesian and spherical coordinates, where the origins of the two spherical angles follow the usual choices for HMDs .

Therefore, the transformation between them is described by the sets of equations:

\begin{equation}
\begin{cases}
\theta = \arctan{\left(x/y\right)}\\
\phi = \arctan{\left(\sqrt{x^2+y^2}~/~z\right)}\\
\end{cases}
\hspace{2mm}
\begin{cases}
x= \sin{\phi} \sin{\theta}\\
y= \sin{\phi} \cos{\theta}\\
z= \cos{\phi}
\end{cases}
\end{equation}

Several planar mappings of the sphere (also called projections) have been considered for the representation of 360 video content: equirectangular, cubemap, pyramidal, equiangular, cubemap...~\cite{el2016streaming,corbillon2017viewport,brown2017bringing}. Since no map of the sphere to the plane can be both conformal and area-preserving, each mapping affects the quality of the different areas of the 360 video content in a different way. Among the different projections that are used, the most common mapping is the equirectangular projection and, therefore, it is the one considered here. Its main advantage lies in the simple transformation equations between the spherical and the planar coordinates. Assuming that the upper-left corner of the frame is the origin of the planar coordinates, the values of $\theta$ and $\phi$ can be scaled directly to obtain their planar counterparts. Therefore, if the 360 video contents are represented in a $N_H \cdot N_V$ frame, the pixel coordinates of point on the sphere $(\theta,\phi)$ are $((\theta/360) N_H, (\phi/180) N_V)$ and, thus, we will keep the $(\theta,\phi)$ addressing for the equirectangular plane. Nevertheless, it must be stressed that any other projection can be considered, as the methodology proposed here can be applied on any geometry.

\subsection{FoV, viewport, and projected viewport}

Let us consider that the FoV is a right rectangular pyramid whose vertex is located in the center of the viewing sphere of the HMD. Thus, the viewport on the sphere is a spherical rectangle centered around the PoG and each one of its four sides is a great circle arc. If the FoV is defined by its two dihedral angles $(\theta_{\text{VP}},\phi_{\text{VP}})$, the solid angle $SA$ subtended at the center of the sphere by the FOV is~\cite{todhunter1886spherical}:

\begin{equation}
SA = 4 \arcsin{\left(\sin\frac{\phi_{\text{VP}}}{2}\sin\frac{\theta_{\text{VP}}}{2}\right)}
\end{equation}

Any value of $(\theta_{\text{VP}},\phi_{\text{VP}})$ is acceptable. However, to help explain the procedure and perform the associated experiments, we have chosen $(\theta_{\text{VP}},\phi_{\text{VP}})~=~(100^o,85^o)$ since they represent the average value of the FoV parameters found in the most common HMDs. Thus, the solid angle is 2.15 steradians, roughly 1/6 of the surface of the sphere, which implies a large deformation, where the four spherical angles are clearly larger than $90^o$. Therefore, linear approximations, commonly employed in the literature, cannot be used.

Although the shape of the projected viewport in the equirectangular image varies significantly according to the location of the viewport on the sphere, let us remember that the size (subtended solid angle) of the viewport is constant. So, it is useful to obtain this constant value in pixel related units. If the sampling rate at the equator of the equirectangular image is taken as reference, each one of the pixels lying on the equator covers a unit area, either in the equirectangular image or on the sphere. However, pixels located outside the equator cover less area on the sphere as the sampling rate along parallels increases with the latitude.

As said before, the origin of coordinates of the equirectangular image is located in the upper-left corner, as shown in Figure~\ref{fig:Corners-of-the-central-FoV}. The area covered by each pixel in the above mentioned area units is $a(\theta,\phi) = \sin{\phi}$. We call the {\it equivalent number of pixels} the area of the region expressed in these area units. Thus, the equivalent number of pixels of the whole frame is $N_{\text{picture}}$:

\begin{equation}
N_{\text{picture}} = \sum\limits_{i.j} a(\theta_i,\phi_j) 
= N_H \sum\limits_{j} \sin{\phi_j} = \frac{2}{\pi} N_H N_V
\end{equation}

As the whole frame covers the whole surface of the sphere, the solid angle subtended is $4 \pi$ steradians. Thus, the equivalent number of pixels of the viewport, $N_{\text{viewport}}$, can be obtained as a proportion of the solid angles subtended:

\begin{equation}
N_{\text{viewport}}  =  \frac{2}{\pi^2} N_H N_V \arcsin{\left(\sin\frac{\phi_{\text{VP}}}{2}\sin\frac{\theta_{\text{VP}}}{2}\right)}
\end{equation}

This result should be rounded if an integer value is required.

Furthermore, this expression of the equivalent number of pixels of the viewport sets the maximum effective resolution that can be achieved by the HMD display. As an example, for the usual values considered in this paper, $(\theta_{\text{VP}},\phi_{\text{VP}})~=~(100^o,85^o)$, $N_H N_V = 3840~\text{x}~1920$, we obtain $N_{\text{viewport}} = 802871$ pixels, clearly lower than 1~Mpixel.

\subsection{Procedure for computing the projected viewport}


Looking for a simpler set of operations to obtain the shape of the projected viewport, we decompose the computation of the projected viewport around the PoG of the user into a three-step procedure. First, we consider a base viewport centered on the central point of the equirectangular image and compute its vertices and several points along its four sides that will help define a piecewise linear approximation of those sides. Then, we rotate this set of points to place them around the PoG of the user. Finally, we obtain the desired projection by connecting those points, thus generating a closed region. Pixels within the boundaries marked by these connections belong to the projection and the set of these pixels is called the mask. These steps are explained in detail below.

\subsection{Base projected viewport}

Every projected viewport is characterized by the location of its four spherical vertices. As stated, we first determine those of the base viewport, which correspond to the initial viewing experience. Throughout the paper and in our experiments, we assume that the user begins looking at the center of the equirectangular image, which corresponds to $O=(180^o,90^o)$. However, the proposed methodology can be adapted to any other desired initial PoG located at the equator, due to the special features of the base viewport that are described below.

\begin{figure}[htbp]
\begin{centering}
\textsf{\includegraphics[width=0.9\columnwidth]{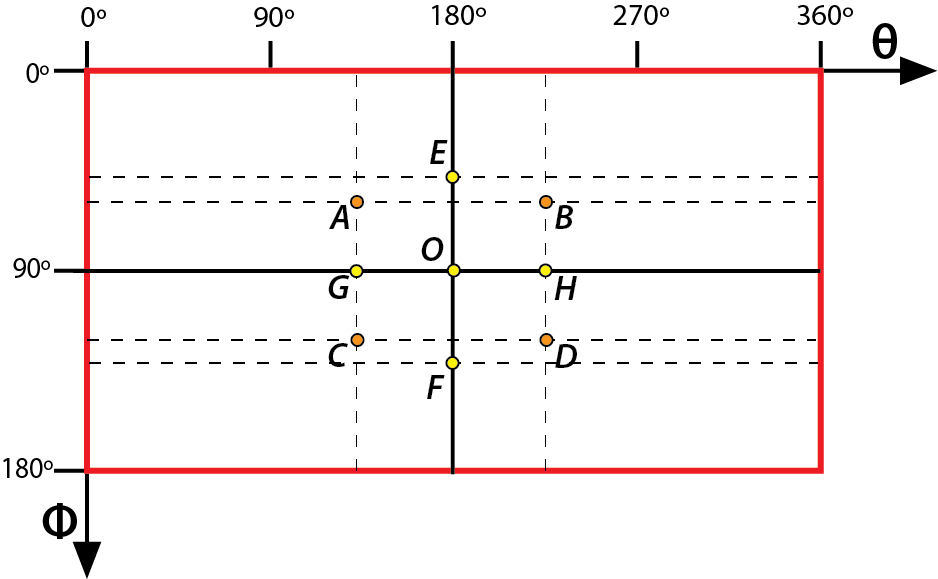}}
\par\end{centering}
\caption{\label{fig:Corners-of-the-central-FoV}{Key points to compute the base viewport}}
\end{figure}

As the base viewport is centered on the equator, the two vertical sides follow two meridians. However, the two horizontal sides do not follow any parallel. The analysis begins determining the coordinates of the middle points of the four sides of the projected viewport ($E$, $F$, $G$ and $H$ in Figure~\ref{fig:Corners-of-the-central-FoV}):


\begin{equation}
\begin{cases}
E = (180^o,90^o-\phi_{\text{VP}}/2)\\
F = (180^o,90^o+\phi_{\text{VP}}/2)\\
G = (180^o-\theta_{\text{VP}}/2,90^o)\\
H = (180^o+\theta_{\text{VP}}/2,90^o)
\end{cases}
\end{equation}

As the viewport sides $AC$ and $BD$ follow two meridians, they are projected as vertical straight lines on the equirectangular image. Therefore their abscissas are equal to those of their midpoints $G$ and $H$ respectively. However, the same does not apply for the other two lines $AB$ and $CD$, requiring the analysis of the projection of the viewport on the sphere to obtain the values of their ordinates:

\begin{equation}
\begin{cases}
\phi_A = \phi_B = 90^o - 2 \arctan\left(\tan{\frac{\phi_{\text{VP}}}{2}}\cos{\frac{\theta_{\text{VP}}}{2}}\right)\\
\phi_C = \phi_D = 90^o + 2 \arctan\left(\tan{\frac{\phi_{\text{VP}}}{2}}\cos{\frac{\theta_{\text{VP}}}{2}}\right)\\
\end{cases}
\end{equation}


Let us now consider the location of the points along the sides. Thus, let $AB$ be a great circle arc, $(x_{A},y_{A},z_{A})$ the Cartesian and $(\theta_{A},\phi_{A})$ the spherical coordinates of point $A$, and $(x_{B},y_{B},z_{B})$ the Cartesian and $(\theta_{B},\phi_{B})$ the spherical coordinates of point $B$. Additionally, let $L$ be another point in the same great circle arc defined by $A$ and $B$ on the unit sphere and let $(x_{L},y_{L},z_{L})$ be its Cartesian and $(\theta_{L},\phi_{L})$ its spherical coordinates. Then, since the three points belong to the same great circle arc, and so to the same plane, the determinant of the matrix built with their Cartesian coordinates is zero:

\begin{equation}
\left|\begin{array}{ccc}
x_{L} & y_{L} & z_{L}\\
x_{A} & y_{A} & z_{A}\\
x_{B} & y_{B} & z_{B}
\end{array}\right|=0,
\end{equation}
Solving the equation and taking into account that the point $L$ lies on the surface of the unit sphere, the equation of the line joining the two vertices $A$ and $B$ is:

\begin{equation}
\phi_{L}=\arctan\left(-~\frac{\gamma}{\alpha\sin\theta_{L}+\beta\cos\theta_{L}}\right),
\end{equation}

where

\begin{equation}
\begin{cases}
{\normalcolor \alpha=} & {\normalcolor \sin\phi_{A}\cos\theta_{A}\cos\phi_{B}-\sin\phi_{B}\cos\theta_{B}\cos\phi_{A}}\\
{\normalcolor \beta=} & {\normalcolor -\sin\phi_{A}\sin\theta_{A}\cos\phi_{B}+\sin\phi_{B}\sin\theta_{B}\cos\phi_{A}}\\
{\normalcolor \gamma=} & {\normalcolor \sin\phi_{A}\sin\theta_{A}\sin\phi_{B}\cos\theta_{B}}\\
{\normalcolor } & {\normalcolor -\sin\phi_{B}\sin\theta_{B}\sin\phi_{A}\cos\theta_{A}}
\end{cases}.
\end{equation}

\begin{figure}[htbp]
\begin{centering}
\textsf{\includegraphics[width=0.45\columnwidth]{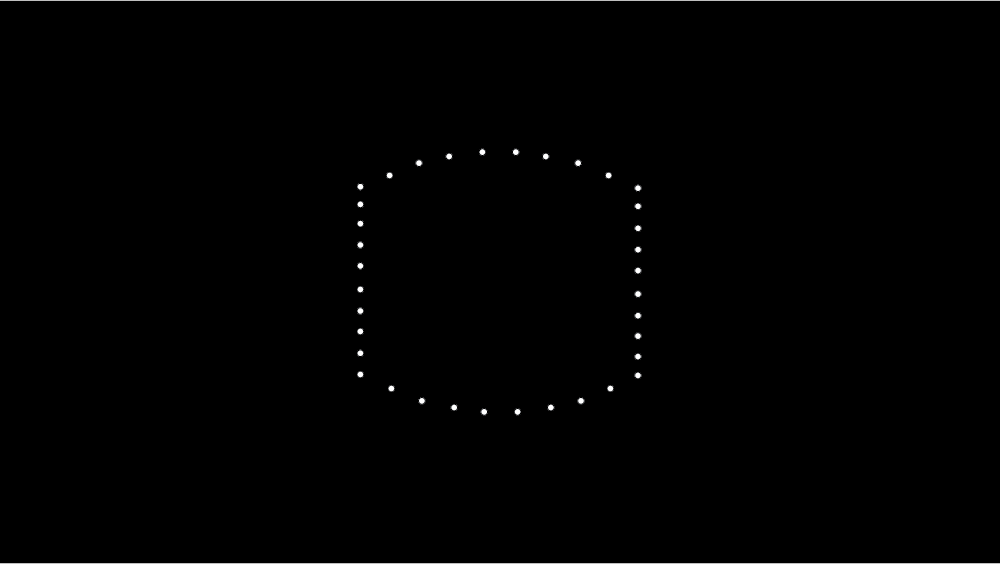}}
\par\end{centering}
\caption{\label{fig:Viewport-projection-after-step1}Base projected viewport (first step of the procedure)}
\end{figure}

We sample the equation for several $\theta_{L}$ values to obtain their corresponding $\phi_{L}$. The resulting $L$ points will be connected later using a piecewise linear function. Therefore, depending on the number of values used, the line approximation will be coarser (low number of points) or more accurate (high number of points). Figure~\ref{fig:Viewport-projection-after-step1} shows the results obtained in this step.

\subsection{Rotation of central viewport}

To simplify the operations in the second step, the viewport is first rotated along $\phi$ and then along $\theta$. Roll movements are not considered, since they are extremely small.

\begin{figure}[htbp]
\begin{centering}
\textsf{\includegraphics[width=4cm]{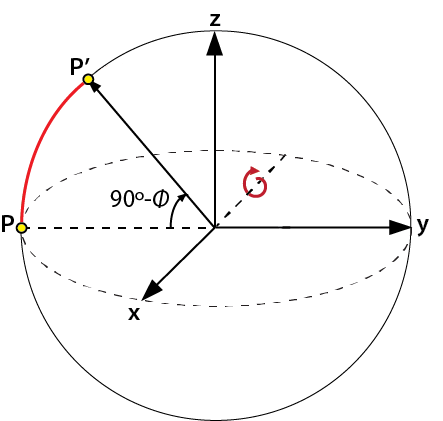}}
\par\end{centering}
\caption{\textcolor{red}{\label{fig:Rotation-about-(-x)}}Rotation about $-x$
axis. Point $P$ is rotated an angle of $(90^o-\phi)$ about
the $-x$ axis, obtaining point $P'$.}
\end{figure}

For the rotation along $\phi$, as the central viewport is so far assumed to be centered in $(0,-1,0)$, it is performed about the $-x$ axis, as can be seen in Figure~\ref{fig:Rotation-about-(-x)}.

Therefore, the corresponding rotation matrix is:

\begin{equation}
\left[\begin{array}{ccc}
1 & 0 & 0\\
0 & \cos(90^o-\phi) & \sin(90^o-\phi)\\
0 & -\sin(90^o-\phi) & \cos(90^o-\phi)
\end{array}\right].
\end{equation}

\begin{figure}[htbp]
\begin{centering}
\textsf{}\subfloat[Before horizontal rotation]{\begin{centering}
\textsf{\includegraphics[width=0.45\columnwidth]{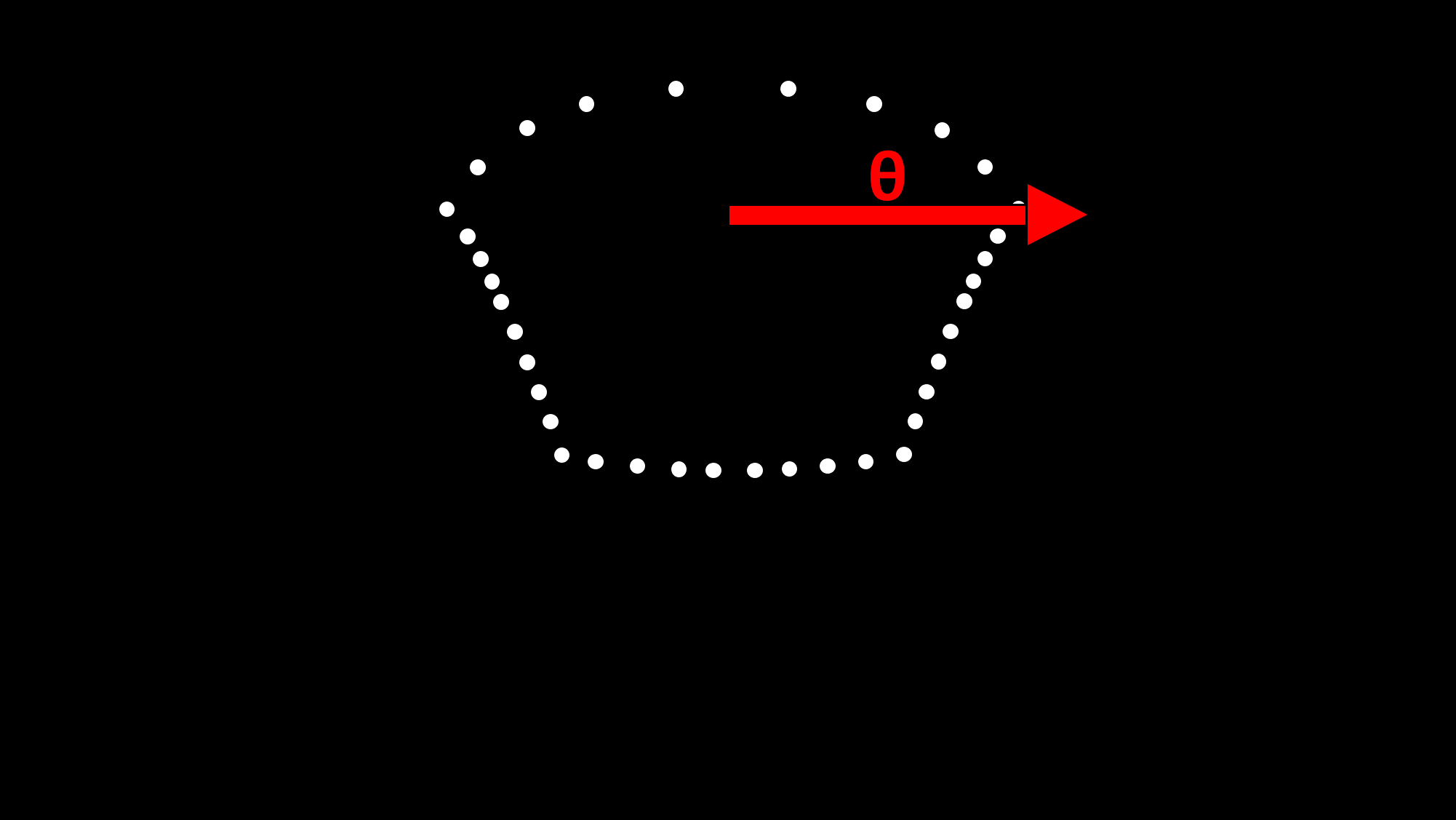}}
\par\end{centering}
\textsf{}}\textsf{}\subfloat[After horizontal rotation]{\begin{centering}
\textsf{\includegraphics[width=0.45\columnwidth]{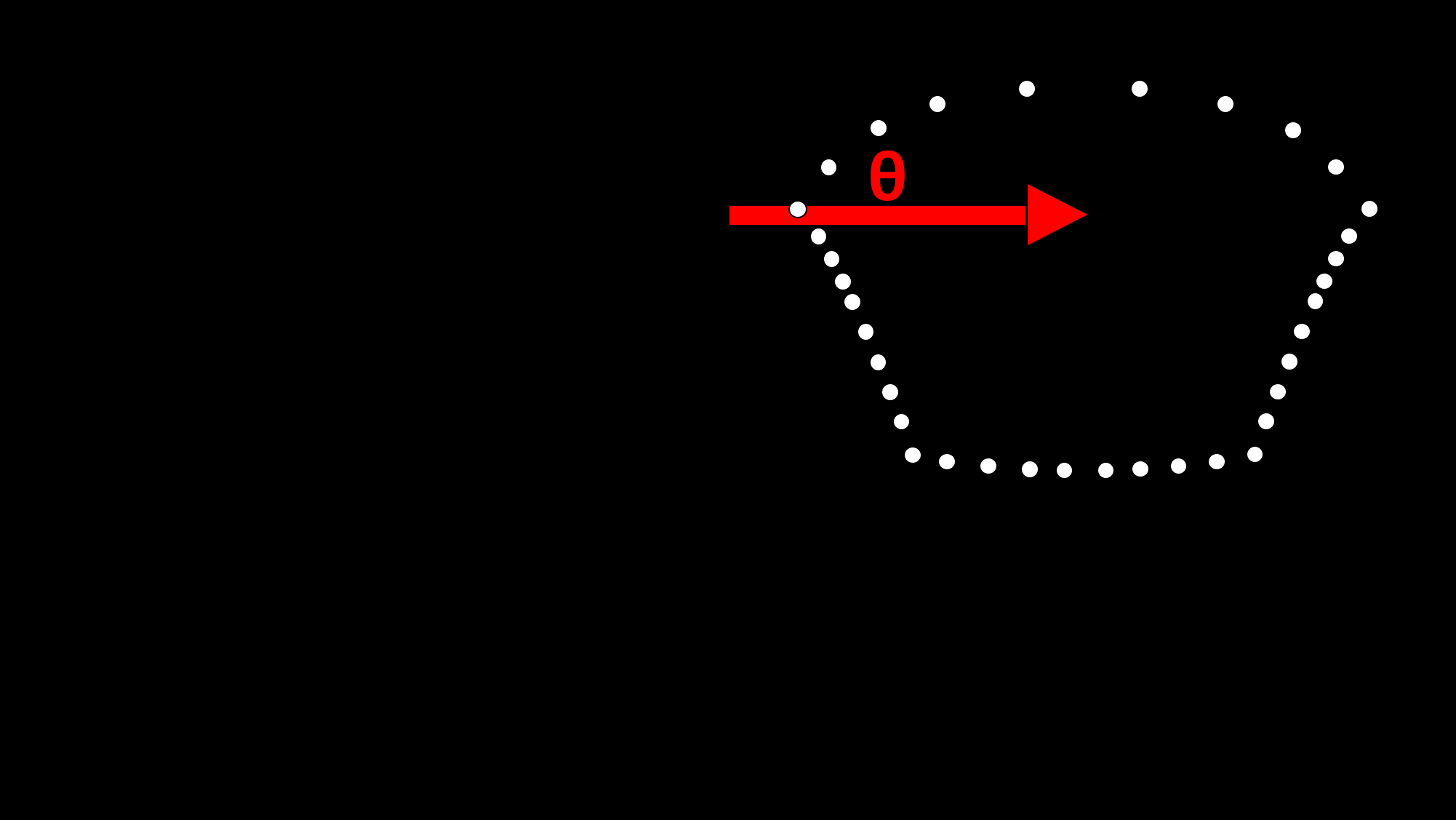}}
\par\end{centering}
\textsf{}}
\par\end{centering}
\caption{\label{fig:RotationTheta}Horizontal rotation of $\theta$ degrees of the projected viewport}
\end{figure}

\begin{figure}[htbp]
\begin{centering}
\textsf{}\subfloat[$\theta=180{^\circ},\phi=30{^\circ}$]{\begin{centering}
\textsf{\includegraphics[width=0.45\columnwidth]{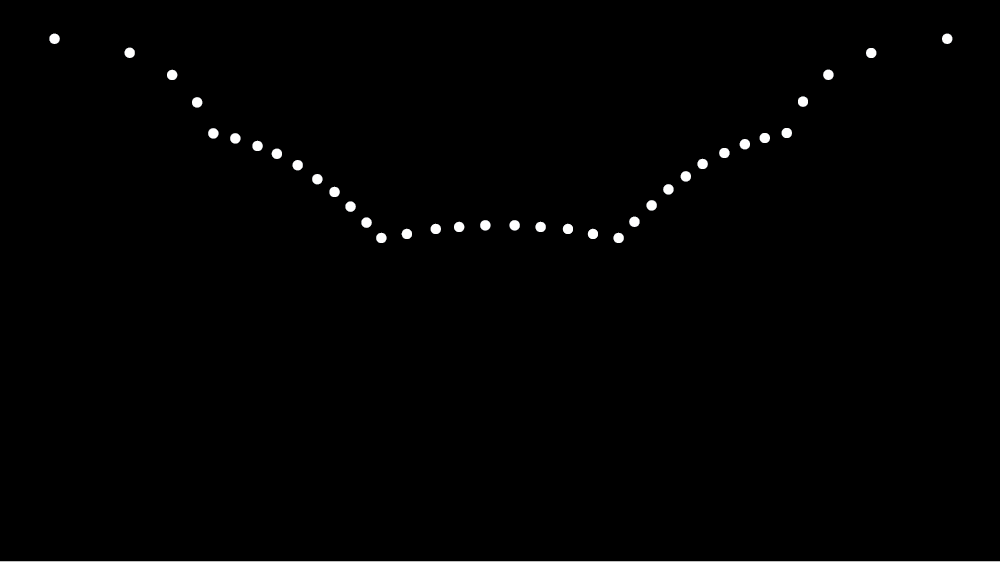}}
\par\end{centering}
\textsf{}}\textsf{}\subfloat[$\theta=320{^\circ},\phi=110{^\circ}$]{\begin{centering}
\textsf{\includegraphics[width=0.45\columnwidth]{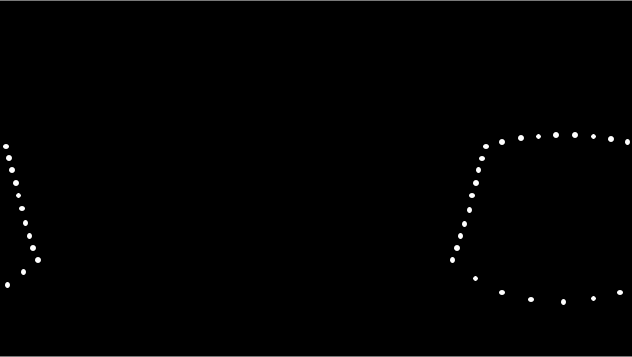}}
\par\end{centering}
\textsf{}}
\par\end{centering}
\caption{\label{fig:RotationExamples}Examples of rotated (vertically plus horizontally) projected viewports}
\end{figure}

The second part of the rotation is performed as follows. Once the viewport has been rotated vertically, it is moved horizontally on the equirectangular image. In Figure~\ref{fig:RotationTheta}, the points of the viewport in (a) are rotated $\theta$ degrees horizontally, obtaining the points shown in (b). More examples of the results at the end of this second step are shown in Figure~\ref{fig:RotationExamples}.

\subsection{Mask generation}

\begin{figure}[htbp]
\begin{centering}
\textsf{}\subfloat[Perimeter using straight lines]{\begin{centering}
\textsf{\includegraphics[width=0.45\columnwidth]{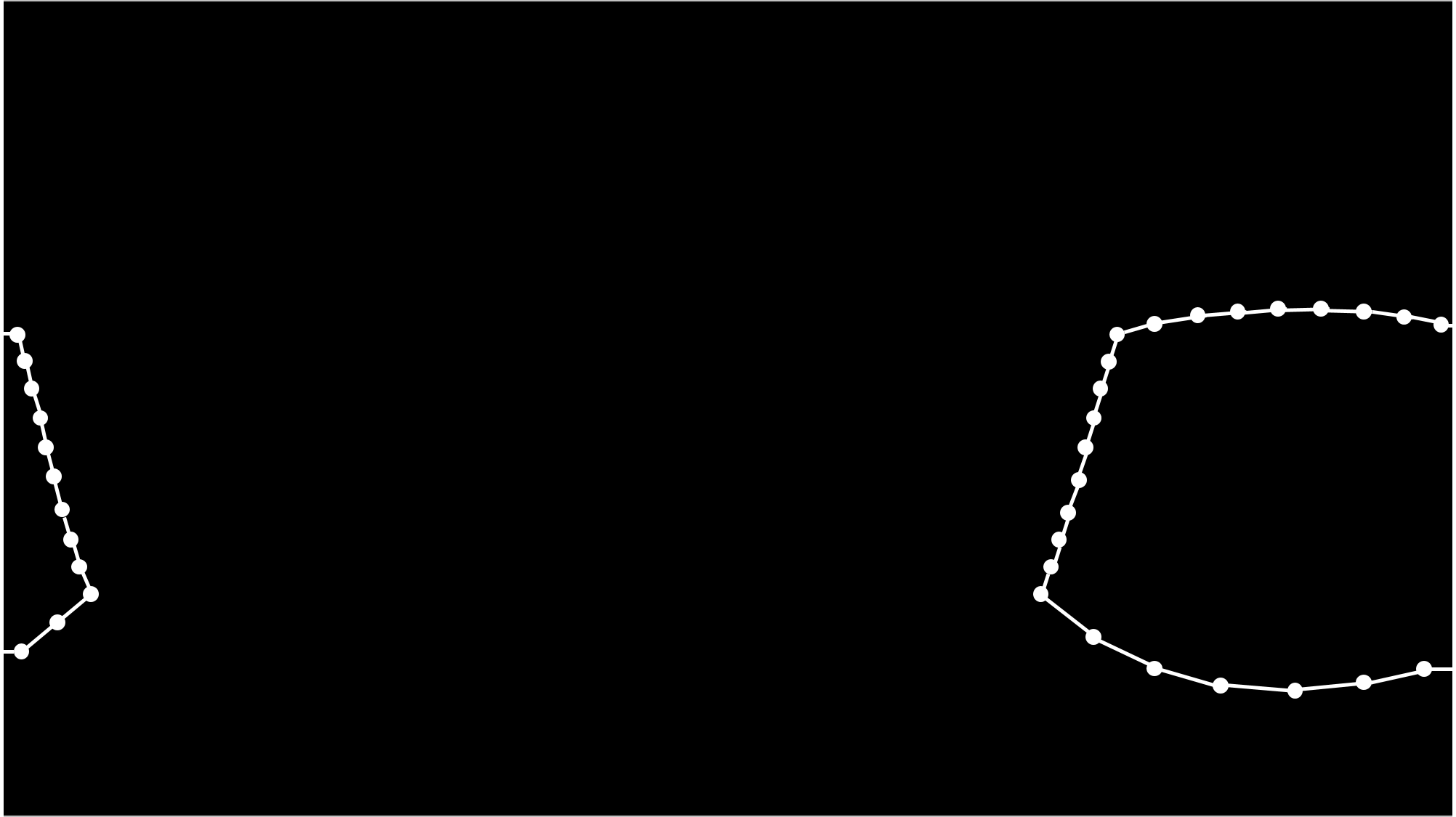}}
\par\end{centering}
\textsf{}}\textsf{}\subfloat[Filled closed region]{\begin{centering}
\textsf{\includegraphics[width=0.45\columnwidth]{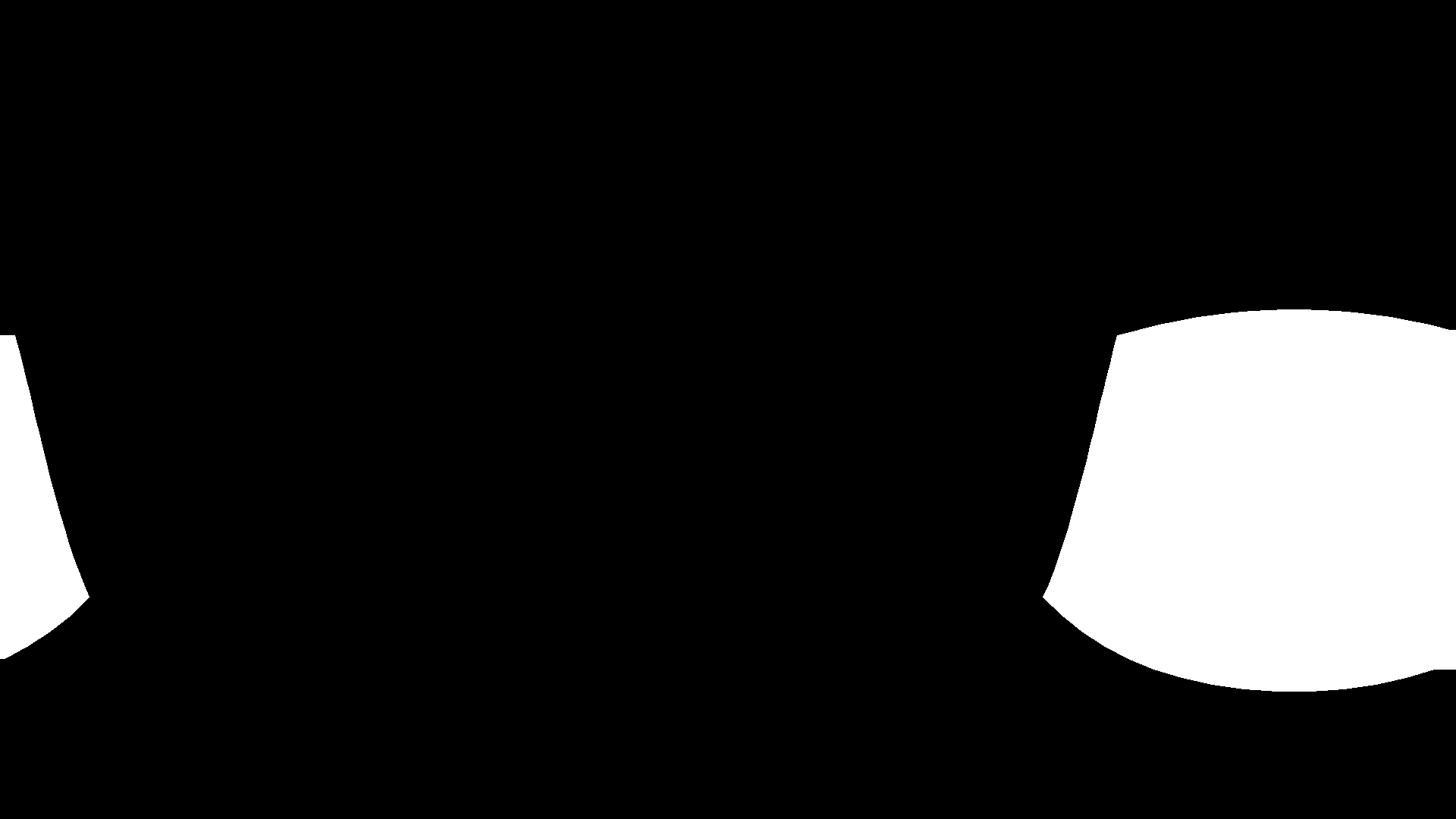}}
\par\end{centering}
\textsf{}}
\par\end{centering}
\caption{\label{fig:Mask-generation}Mask generation procedure (last step of the procedure)}
\end{figure}

\begin{figure}[htbp]
\begin{centering}
\textsf{}\subfloat[$\theta~=~320^o$, $\phi~=~110^o$]{\begin{centering}
\textsf{\includegraphics[width=0.45\columnwidth]{mask_phi110_theta320}}
\par\end{centering}
\textsf{}}\textsf{}\subfloat[$\theta~=~180^o$, $\phi~=~30^o$]{\begin{centering}
\textsf{\includegraphics[width=0.45\columnwidth]{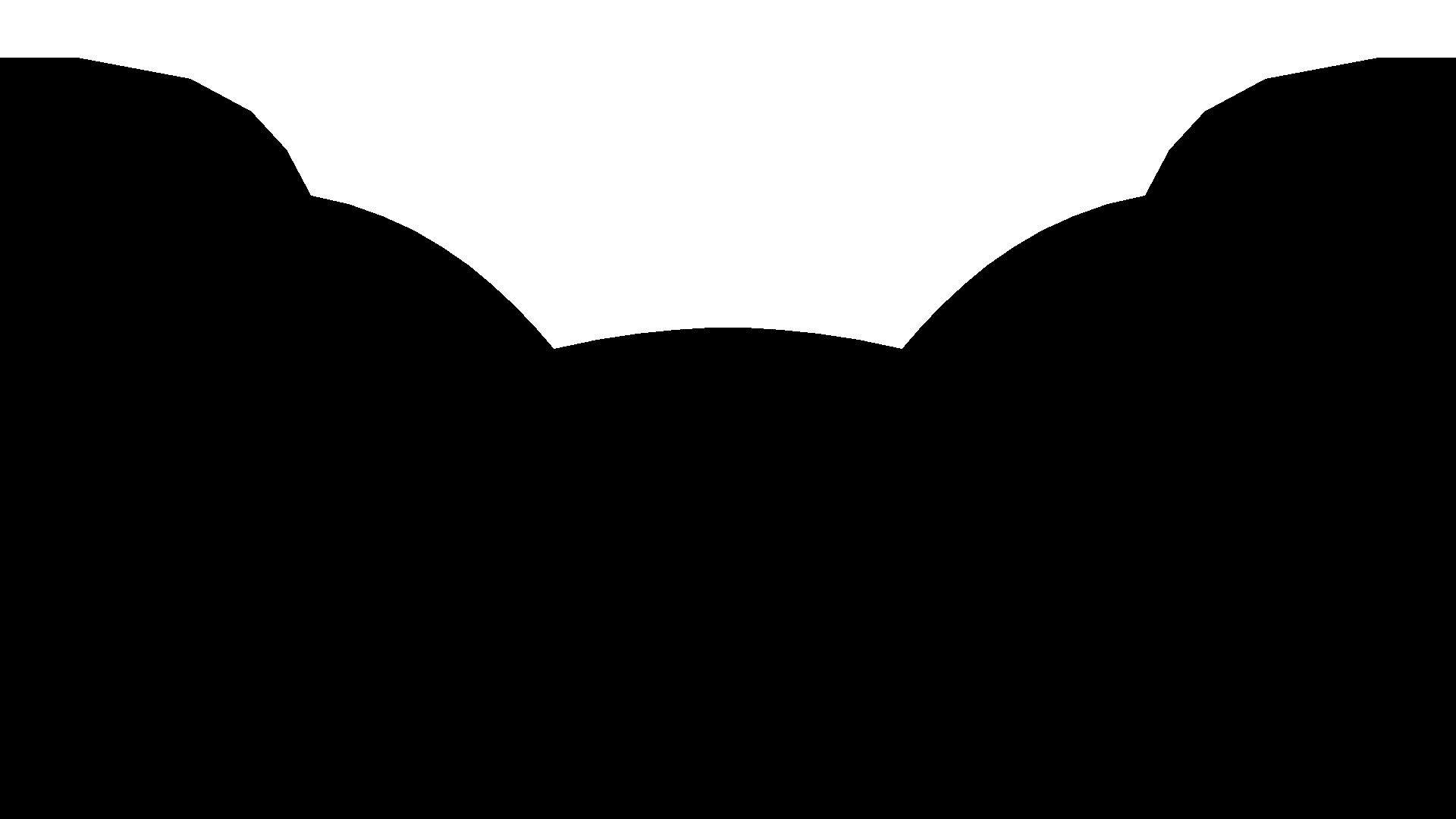}}
\par\end{centering}
\textsf{}}
\par\end{centering}
\begin{centering}
\textsf{}\subfloat[$\theta~=~180^o$, $\phi~=~180^o$]{\begin{centering}
\textsf{\includegraphics[width=0.45\columnwidth]{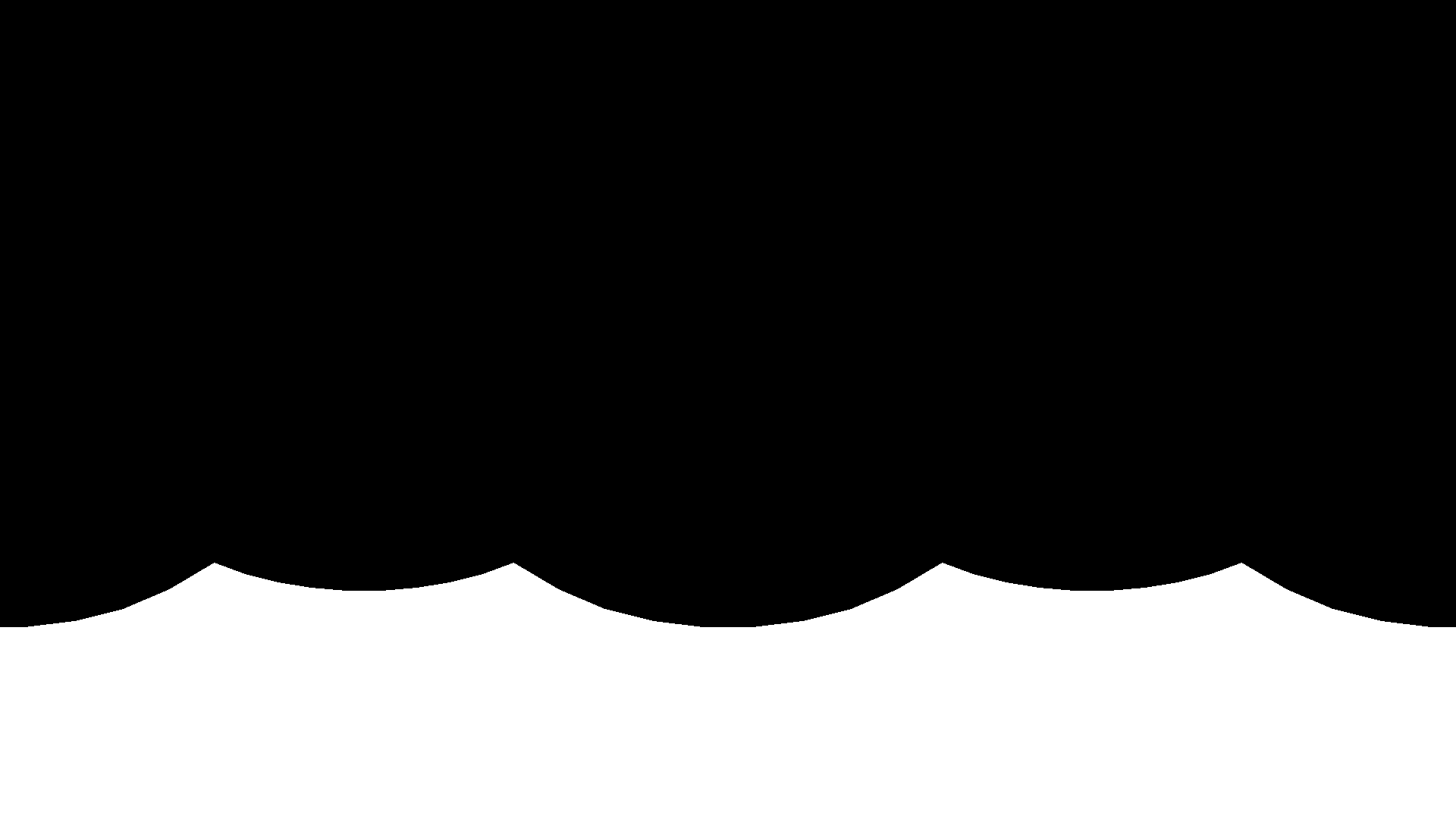}}
\par\end{centering}
\textsf{}}\textsf{}\subfloat[$\theta~=~180^o$, $\phi~=~60^o$]{\begin{centering}
\textsf{\includegraphics[width=0.45\columnwidth]{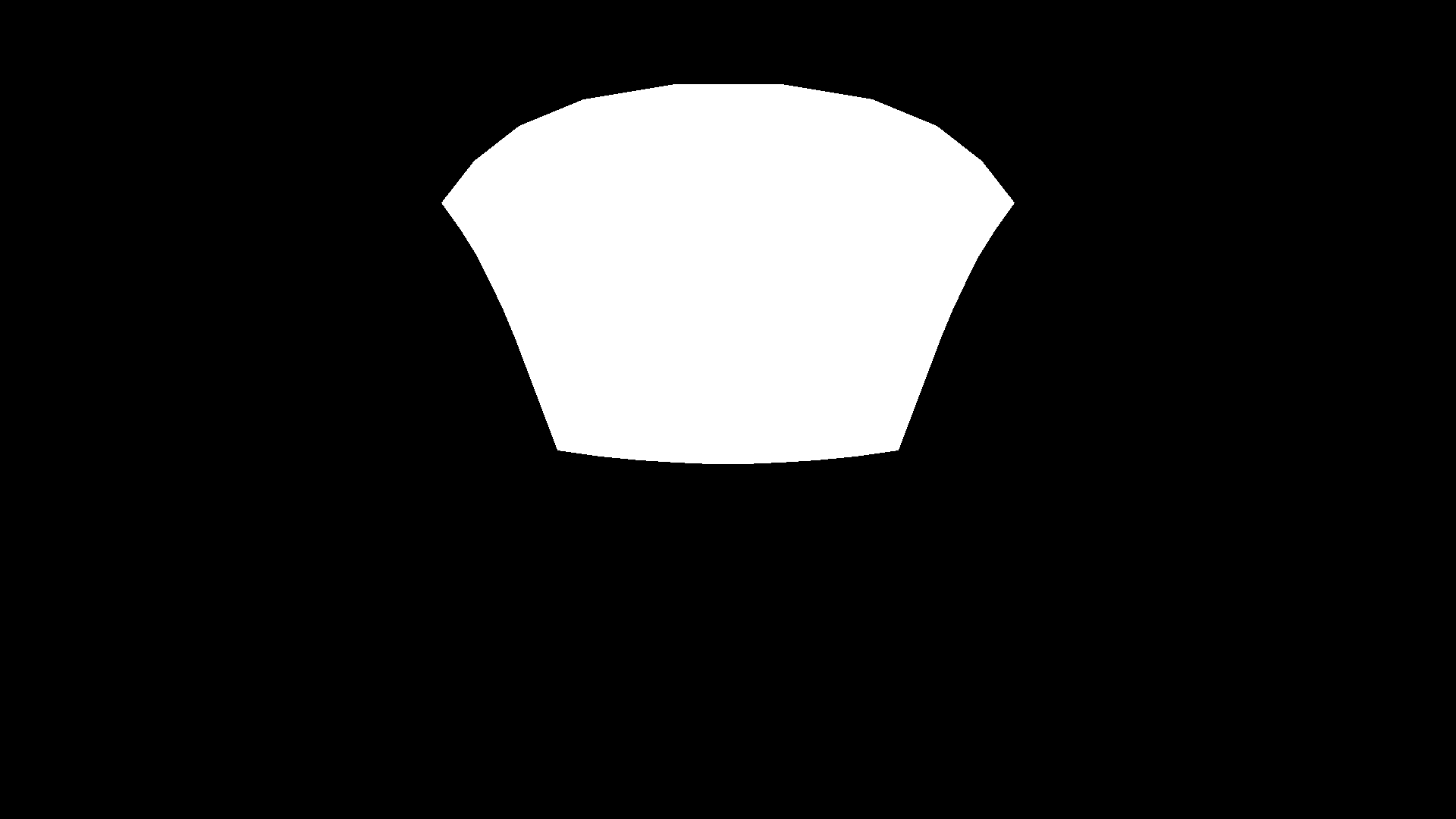}}
\par\end{centering}
\textsf{}}
\par\end{centering}
\caption{{\label{fig:Examples}}Examples of masks centered at different locations}
\end{figure}

After the first two steps, the obtained points are joined with straight lines, that is, using a piecewise linear function, generating a closed region. This closed region is then filled to obtain the desired mask. Different possible situations must taken into account to correctly identify the region of the image within the mask. For example, Figure~\ref{fig:Mask-generation} presents an special where the mask is not totally connected, but divided in two parts due to circular shifts. Additional examples of the obtained masks are shown in Figure~\ref{fig:Examples}.

At this point, we have a binary mask, $M_{i}$, where the non-zero elements represent the viewport projection on frame $i$. However, to compensate for the unequal sampling of the sphere by the equirectangular projection, the values of the elements in $M_{i}$ are weighted according to the area they cover on the sphere. Each of these weights depends exclusively on the latitude and its value is equal to the sine of its corresponding $\phi$ value ($w^{a}_{j} = \text{sin}~\phi_{j}$). 

\begin{equation}
    \tilde{m}_{i,p} = m_{i,p} \cdot w^{a}_{p} ~~~~~ \forall p \in M_{i} 
\end{equation}

Furthermore, the values of the corresponding pixels can be weighted additionally to provide more importance to the more relevant pixels within the viewport, i.e. the central area with respect to the viewport edges. In this case, the pixels in the viewport are weighted considering the distance to the PoG $w^{c}_{j} = f\left( d\left(j,j_{\text{PoG}} \right)\right)$ and normalized accordingly.






\section{\label{sec:Methodology}Proposed Methodology}

\begin{figure}[htbp]
\begin{centering}
\textsf{\includegraphics[width=0.9\columnwidth]{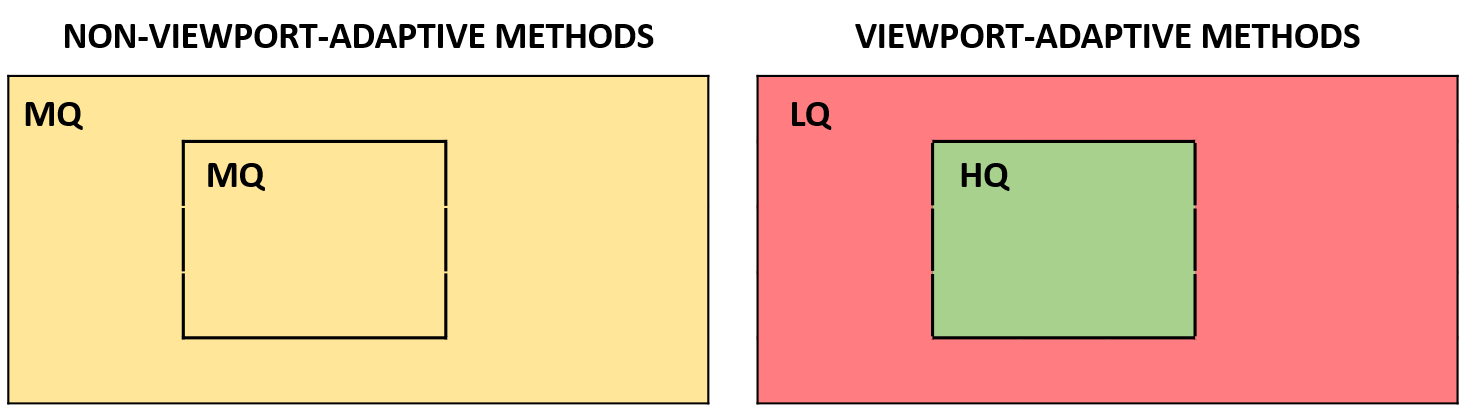}}
\par\end{centering}
\caption{\label{fig:Example-of-a-quality-matrix} Difference between non-viewport-adaptive and viewport-adaptive methods regarding the quality distribution (LQ: Low Quality, MQ: Medium Quality and HQ: High Quality)}
\end{figure}

Viewport adaptation implies that the quality is not uniformly distributed on the image, but is composed of areas of different qualities. Figure~\ref{fig:Example-of-a-quality-matrix} shows the difference between both non-viewport-adaptive and viewport-adaptive methods, where LQ, MQ and HQ correspond to low quality, medium quality and high quality, respectively. Although the figure presents only two qualities for the non-viewport-adaptive method, more qualities can be used as long as the bitrate is preserved. This non-uniform quality distribution enables that users may observe more than one quality at the same time. The information about the corresponding quality of each of the areas of the image is represented by a grade matrix $V$, where each entry represents the quality value of one pixel of the image. As mentioned in the introduction, these non-uniform distributions can be implemented thanks to the use of tiles, since each of them may have a different quality.

\begin{figure}[t]
\begin{centering}
\textsf{\includegraphics[width=0.9\columnwidth]{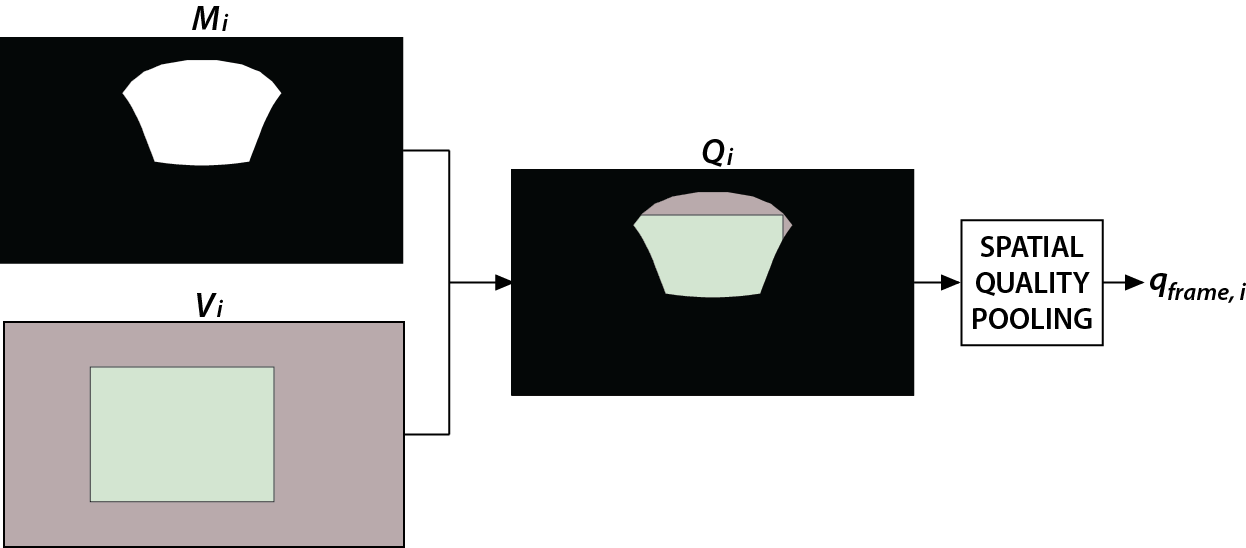}}
\par\end{centering}
\caption{\label{fig:Quality-metric}Quality metric procedure to obtain the quality value for frame $i$}
\end{figure}

The main idea of the proposed methodology is to provide a figure of merit that reflects the quality really perceived by the user along a temporal window. Therefore, we need a representative value of the quality seen at each frame within this temporal scope. To that end, we define the quality $q_{i,p}$ of each pixel $p$ as the product of a geometric-related component, $m_{i,p}$, and a grade-related one, $v_{i,p}$. The matrix $M_{i}$, outcome of the previous section as the mask representing the viewport projection, contains the geometric-related component of each pixel at frame $i$. A second matrix, $V_{i}$, contains the grade-related components of each pixel of the image of the viewport-adaptive content presented to the user at that particular moment. Thus, the resulting matrix $Q_{i}$ can be formulated as the Hadamard product of the geometric-related and the grade-related matrices:

\begin{equation}
Q_{i}=M_{i} \circ V_{i}.
\end{equation}

A spatial quality pooling can be obtained for every frame and a temporal quality pooling can be computed to obtain an overall quality figure for the considered temporal window.


%


Regarding the spatial pooling, the quality value $q_{\text{frame}, i}$ for frame $i$ is computed as the average of all the quality values within $Q_{i}$ of the pixels in the viewport projection,

\begin{equation}
q_{\text{frame}, i}=\frac{1}{N_{\text{viewport}}}\sum_{p\in M_{i}} q_{i,p}
\end{equation}
where $N_{\text{viewport}}$ is the equivalent number of pixels of the viewport obtained in the previous section, and $q_{i,p}$ is the element in matrix $Q_{i}$ representing the quality of pixel $p$. Since all the values outside the viewport projection are null, the summation can be extended to the whole image, delivering the same value:
\begin{equation}
q_{\text{frame}, i}=\frac{1}{N_{\text{viewport}}}\sum_{p\in Q_{i}} q_{i,p}
\end{equation}

The proposed methodology up to the output of the spatial quality pooling is illustrated in Figure~\ref{fig:Quality-metric}.

Regarding the temporal pooling, we have defined two different approaches: the mean and the fraction of time above a threshold. The first one reflects the average quality shown to the user during a temporal window of the streaming session and is obtained by a uniform or weighted average of the spatial quality over time, leading to $q_{\text{window}}$,

\begin{equation}
q_{\text{window}}=\frac{1}{N_{f}} \sum_{i=0}^{N_{f}-1} q_{\text{frame}, i}
\end{equation}
where $N_{f}$ is the number of frames in the analyzed temporal window. The second approach gives a figure of the fulfillment of a minimum spatial quality along the analyzed window and is computed as the percentage of frames with a quality value higher than a threshold $T_\text{Q}$:


\begin{equation}
f_{\text{window}} =\frac{1}{N_{f}}\sum_{i=0}^{N_{f}-1} \left[q_{\text{frame}, i}~\mbox{>}~T_\text{Q}\right]
\end{equation}
where $\left[P\right]$ is the Iverson bracket, i.e. 1 if $P$ is true and 0 otherwise. The threshold $T_\text{Q}$ can be set to any specific value between 0 and 1 that designers decide better suited to their goals. The greater $T_\text{Q}$ is, the more strict the imposed requirements are in terms of maintained quality over time, taking into account the specifics of the session (type of content, user behavior, setup\dots).

There are two approaches for the quality pooling. On the one hand, if the objective is to evaluate the proportion of high quality area that is presented to the user along the considered temporal window, the entries of matrix $V_{i}$ must be set either to one, if they belong to the high quality area, or to zero, otherwise. On the other hand, if the objective is to look for an objective assessment, any metric that provides a value per pixel can be used for populating matrix $V_{i}$.

\begin{figure}[htbp]
\begin{centering}
\textsf{\includegraphics[width=0.9\columnwidth]{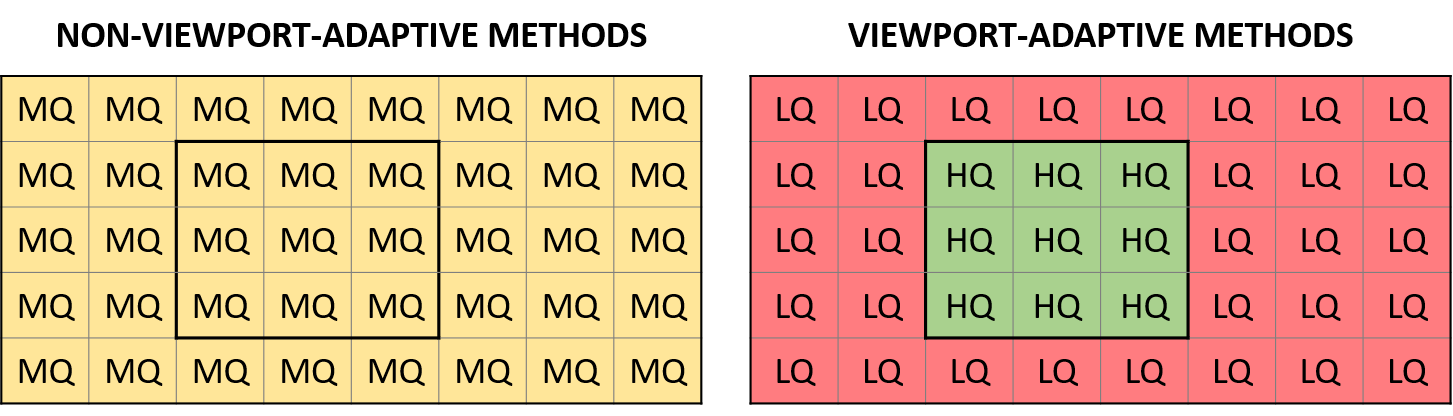}}
\par\end{centering}
\caption{\label{fig:Example-of-a-quality-matrix-area} Difference between non-viewport-adaptive and viewport-adaptive methods regarding the quality distribution (LQ: Low Quality, MQ: Medium Quality and HQ: High Quality)}
\end{figure}

Additionally, non-pixel-based objective metrics such MSE, PSNR, SSIM or VMAF, which provide a single value per frame, can be used by computing $V_{i}$ on a per area basis. In this case, we can exploit the fact that the equirectangular image can be divided into tiles for encoding and therefore we can apply the desired technique to each of them individually or to sets of them (Figure \ref{fig:Example-of-a-quality-matrix-area}). In this way, all the pixels belonging to the same tile or set of tiles will have the same value in the grade matrix. Nevertheless, although the matrix $V_{i}$ is computed in a different way, the proposed methodology is maintained.

\section{\label{sec:Methods}Methods}
The methodology presented in the previous section requires the computation of matrices $M_{i}$ and $V_{i}$ for all the frames in the session. However, computing requirements might be a burden for lightweight real-time applications. Thus, we have defined two methods: the full method, called Viewport Adaptive Quality Method (VAQM), and a lighter version, called Approximated Viewport Adaptive Quality Method (AVAQM), where matrices $M_{i}$ and $V_{i}$ are selected from pre-computed sets of masks and grades to speed up the process. Both methods are described next.

\begin{figure}[htbp]
\begin{centering}
\textsf{\includegraphics[width=0.9\columnwidth]{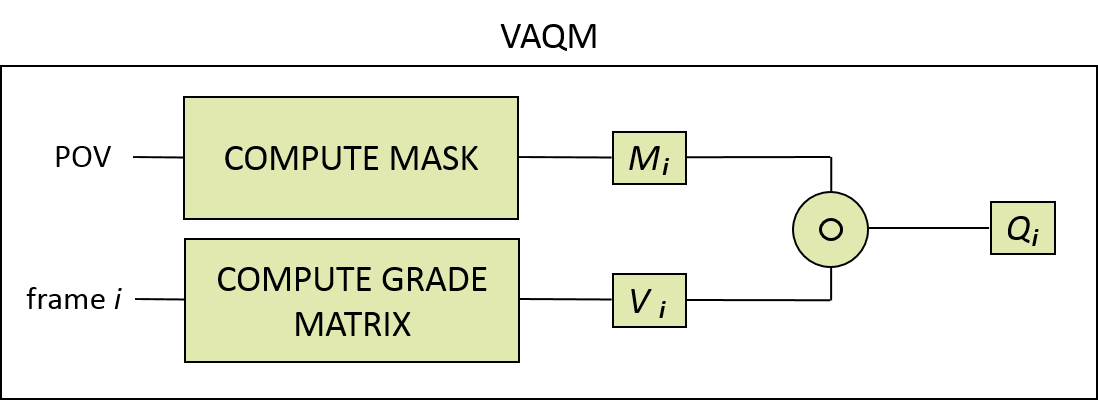}}
\par\end{centering}
\caption{\label{fig:Method}Scheme followed by VAQM}
\end{figure}

\subsection{VAQM (Viewport Adaptive Quality Method)}
The scheme followed in the full method is shown in Figure~\ref{fig:Method}. During the session, we constantly collect information about the content that the user is watching and the PoG of the user at each moment. Afterwards, the viewport projection is computed for all the collected samples, obtaining a new mask for each instant of time. Moreover, the matrix $V_{i}$ is generated from applying the desired quality metric (FR or not) to the whole image (e.g. MSE, PSNR, SSIM or VMAF). Thus, this matrix is of the same size as the transmitted video images. In summary, with this approach, we obtain a very high accuracy in exchange for a greater computational cost.

\subsection{AVAQM (Approximated Viewport Adaptive Quality Method)}
\begin{figure}[htbp]
\begin{centering}
\textsf{\includegraphics[width=0.9\columnwidth]{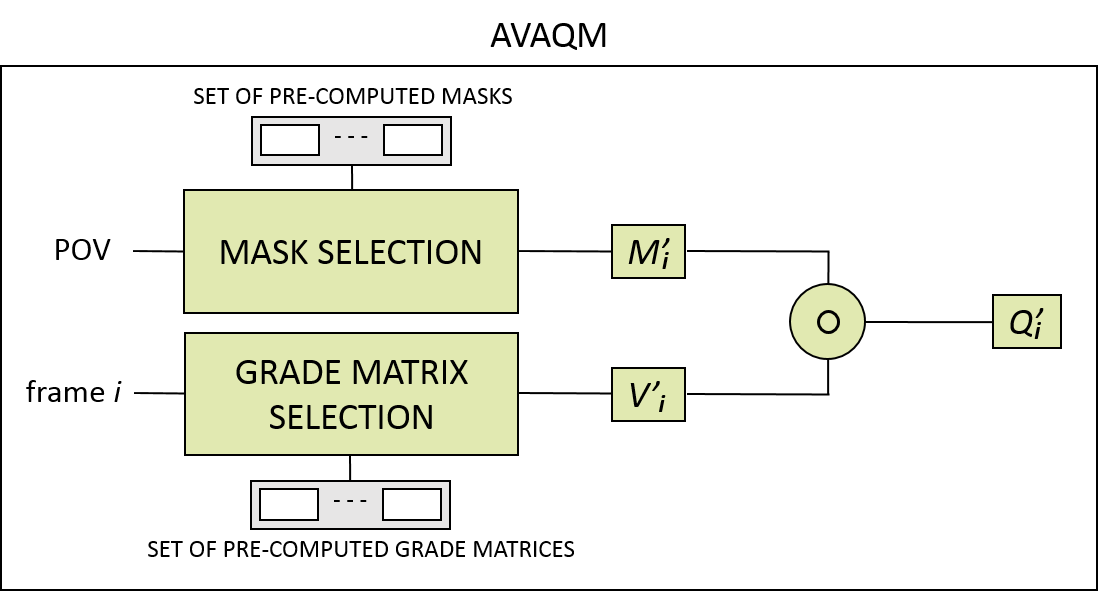}}
\par\end{centering}
\caption{\label{fig:Method-approximation}Scheme followed by AVAQM}
\end{figure}

This approach arises from the fact that there are some scenarios where time restrictions may not allow us to apply the previous method directly, as, depending on the resources, it could be computationally costly. Additionally, it is also valid for when accuracy requirements are more flexible. The scheme followed by this approach is shown in Figure~\ref{fig:Method-approximation}. 

Approximations might be carried out independently on two fronts: in the geometric-related part of the procedure to generate matrix $M_{i}$, and in the grade-related part to obtain $V_{i}$.

\begin{figure}[htbp]
\begin{centering}
\textsf{\includegraphics[width=0.9\columnwidth]{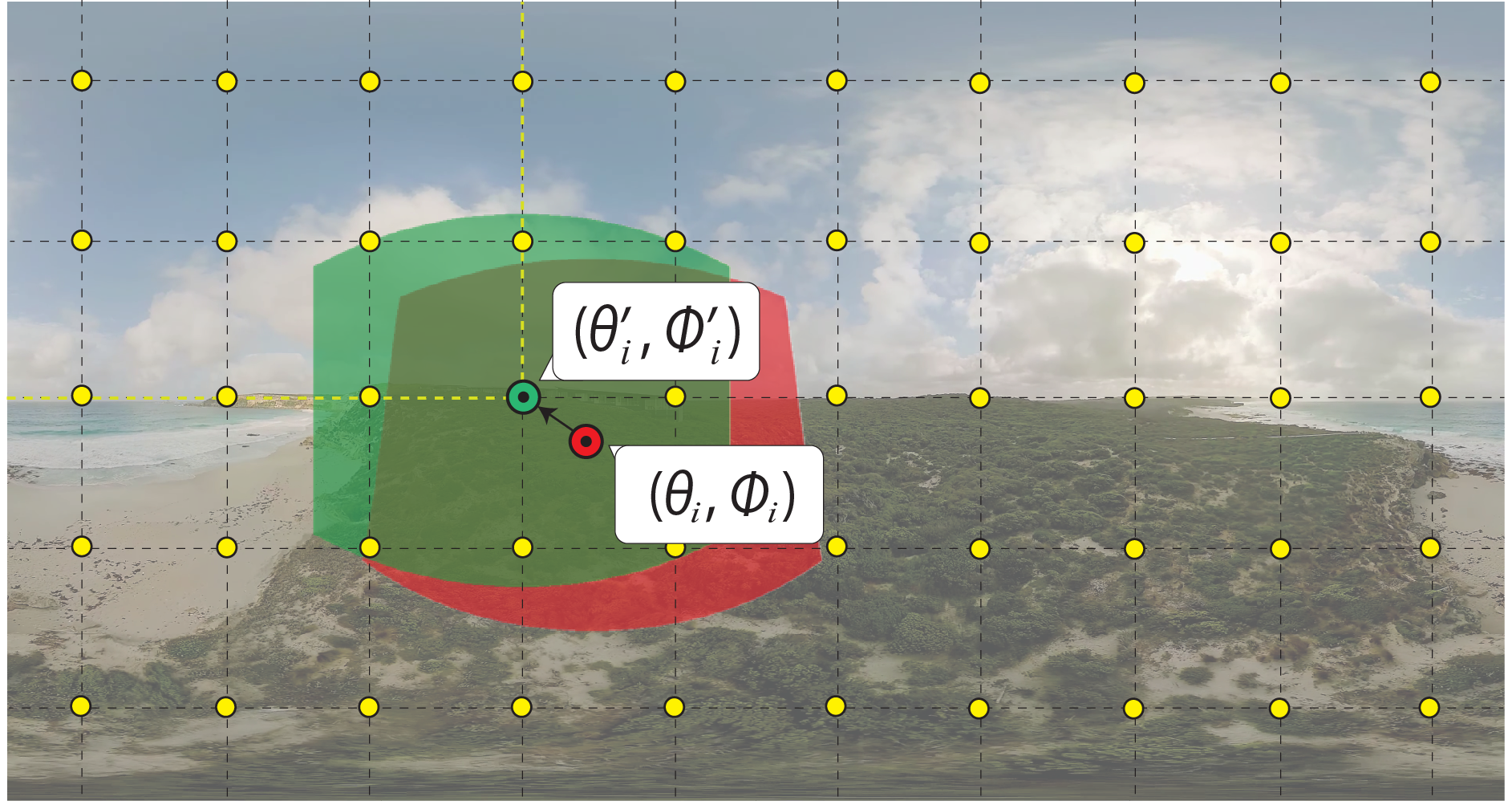}}
\par\end{centering}
\caption{\label{fig:setOfMasks}Example of approximation of matrix $M_{i}$ using 5x10 pre-calculated masks}
\end{figure}

Regarding the geometric part, $M_{i}$ can be approximated using a finite set of pre-calculated masks with centers uniformly distributed throughout the equirectangular image, as represented by the yellow circles in Figure~\ref{fig:setOfMasks}. The selected pre-calculated mask for frame $i$, $M'_{i}$, (in green in the same figure) is the one whose center $\left(\theta'_{i}, \phi'_{i}\right)$ is the nearest to the projected PoG of the user $\left(\theta_{i}, \phi_{i}\right)$ (in red in the same figure). On the plus side, the use of the geometric approximation hugely decreases the computation load required to perform the viewport projection. On the down side, there is a certain loss of accuracy and a slight increase of storage requirements. Both drawbacks heavily depend on the number of pre-calculated masks used.

With respect to matrix $V_{i}$, the entries of its approximation $V'_{i}$ can be computed considering the value of encoding parameters that do not express the resulting image quality but indirectly provide a sufficiently accurate idea of it, like the Quantization Parameter (QP) used to encode the basic processing units (e.g. Coding Tree Units -CTUs- in H.265/HEVC) in the image. In this particular case, lower values correspond to better qualities. The advantage of this approximation relies on the ease and speed when obtaining and mapping such values. The drawback is the gap between these values and any others resulting from the application of a given quality metric in terms of capacity upon representing the quality perceived by users.

Finally, the approximated matrix $Q'_{i}$ is computed in an analogous way as before as the Hadamard product of matrices $M'_{i}$ and $V'_{i}$:

\begin{equation}
Q'_{i}=M'_{i} \circ V'_{i}.
\end{equation}

We include in Table~\ref{tab:Mean-error-Mask-aprox} the mean relative error that results from using different numbers of pre-calculated masks. The study has been carried out with videos of resolution UHD-4K (3840x1920p). Moreover, the QP value used to encode a set of pixels is used as the quality value of these pixels. Nevertheless, this last feature has no real impact on the results, since, as mentioned, the study aims at measuring and registering relative values. To help understand these values, please remember that the viewport covered around 1/6 of the surface of the sphere.

\begin{table}[htbp]
\centering
\caption{\label{tab:Mean-error-Mask-aprox}Mean relative error for different geometric approximations}

\begin{tabular}{c|c}
Number of masks & Mean relative error\\
\hline 
3x6 & 3.78\%\\
5x10 & 2.16\%\\
10x20 & 0.69\%\\
20x40 & 0.29\%
\end{tabular}
\end{table}

Based on the results shown in the table, it can be concluded that the approximation with 10x20 masks constitutes a good trade-off between accuracy (mean relative error of less than 1\%) and storage requirements.

\section{\label{sec:Results}Experiments and results}
The methodology explained in the previous section has been tested through two main sets of experiments. The first one considers the effect of the length of the video segments, whereas the second one is focused on the impact of the movements of the user. Before describing them in depth, we present the test features that are common to them.

\begin{figure}[t!]
\centering
\includegraphics[width=0.9\columnwidth]{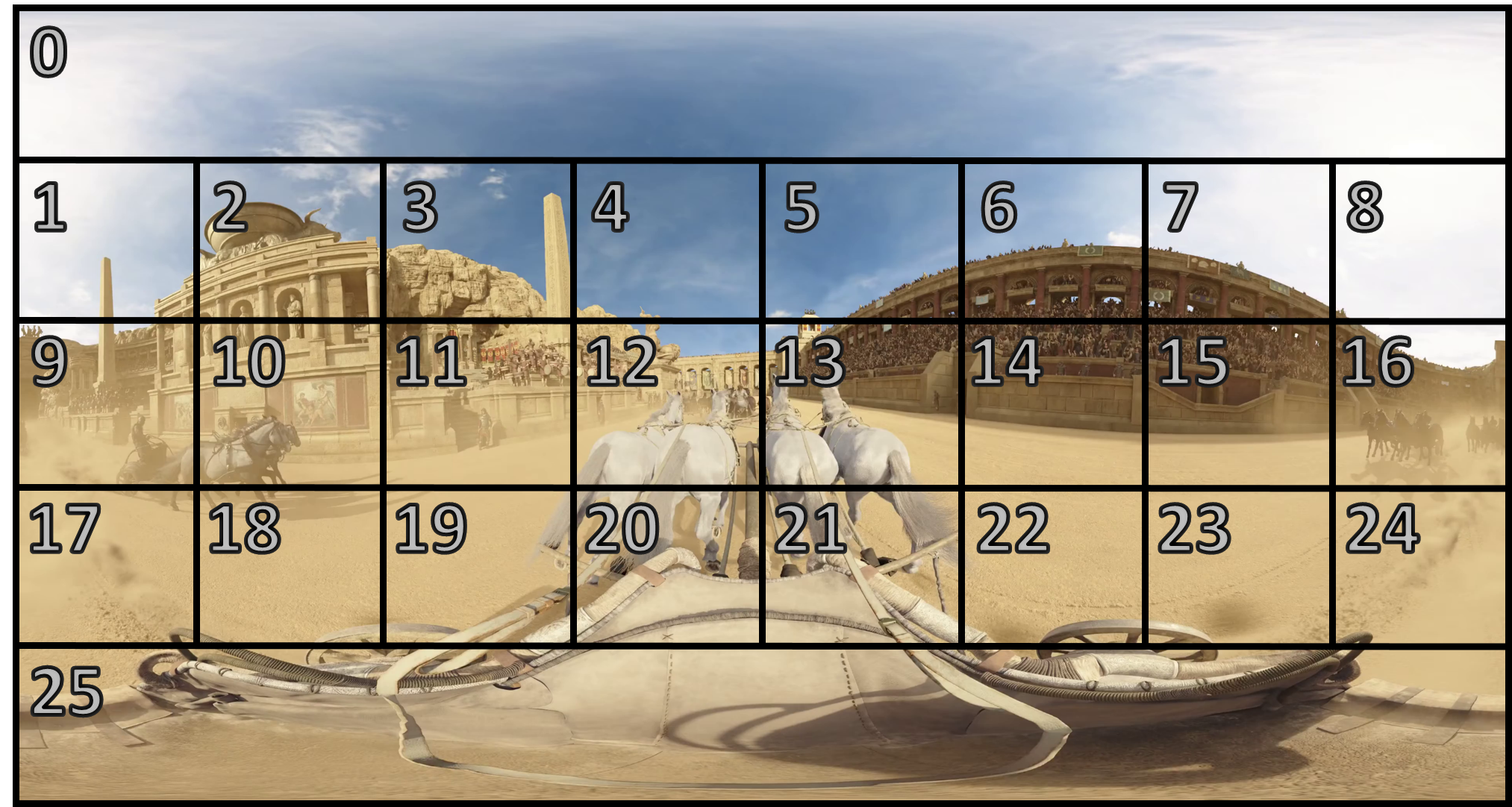}
\caption{\label{fig:Areas 26}Areas in the 360 image associated to the generated viewport-oriented sequences}
\end{figure}

\begin{figure}[t]
\centering
\subfloat[]{\includegraphics[width=0.48\columnwidth]{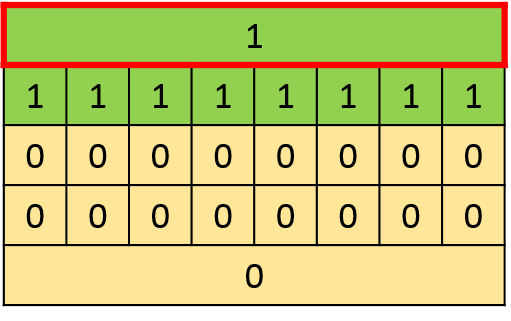}}~
\subfloat[]{\includegraphics[width=0.48\columnwidth]{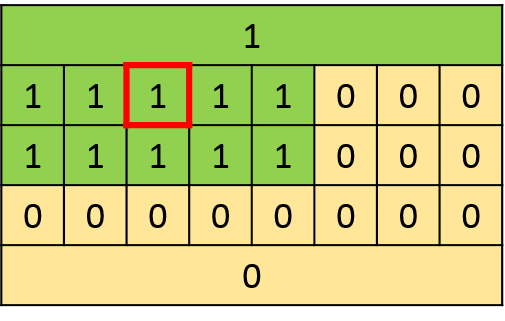}}\\
\subfloat[]{\includegraphics[width=0.48\columnwidth]{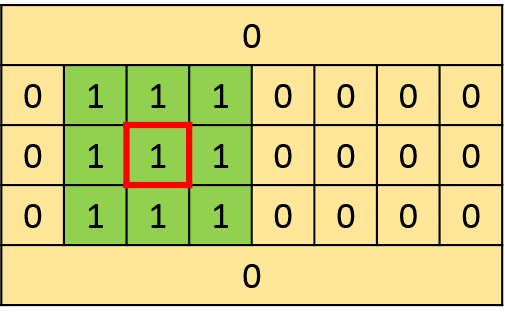}}~
\subfloat[]{\includegraphics[width=0.48\columnwidth]{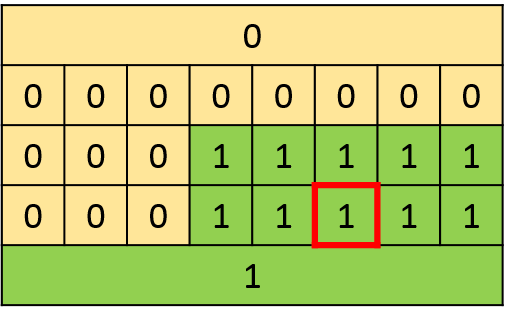}}
\caption{\label{fig:Quality matrices}Distribution of qualities corresponding
to four different areas. The corresponding area is boxed in red.}
\end{figure}

The videos used in the experiments come from the public database 360 Video Viewing Dataset in Head-Mounted Virtual Reality~\cite{lo2017360}. This database includes 10~one-minute-long videos of 3840x1920~pixels at 30~fps. For each source, it also includes the trajectory followed by 50~different users at one sample per frame. Regarding the content preparation, the sequences were H.265/HEVC encoded with 5x8~tiles, a trade-off between coding efficiency and sufficient granularity to generate efficient viewport-adaptive content. Each viewport-oriented sequence, that is, the version of the content created to be presented to the user when his/her PoG lies in a specific area of the sphere, is the result of encoding the source content with a given distribution of qualities. Each of these sequences is associated with one of the non-overlapping areas in the equirectangular image depicted in Figure~\ref{fig:Areas 26}, where the upper and lower stripes of tiles have been merged into one area, respectively. Thus, we have generated 26~viewport-oriented sequences per video source, which are later on segmented to be used in an ABR platform. Additionally, we assume that virtually the whole motion-to-photon latency of the system corresponds to the time that it takes the segment currently in the process of decoding and presentation (which corresponded to the previous PoG) so that the one that is correctly adapted to the PoG can start the same process. Thus, we assume that the time required to download segments is negligible. Furthermore, as a consequence, there are no stalls.

Regarding the configuration related with the proposed methodology, the experiments have been performed using the approximated version of the methodology, that is, with pre-calculated masks for the viewport projections and a set of pre-calculated qualities for each of the areas in the image. As said before, we have assumed a FoV of 100$^o$~horizontally and 85$^o$~vertically. Based on the results shown in Table~\ref{tab:Mean-error-Mask-aprox}, we use 10x20 pre-calculated masks. Finally, for the set of pre-calculated qualities, we have considered the use of the values 1 and 0 for the high and low quality, respectively. Four examples of the distribution of both values in the areas in as many viewport-oriented sequences is depicted in Figure~\ref{fig:Quality matrices}.

Finally, the timeline considered for each session is that of the duration of the sequence presented to the user. Therefore, the window for the temporal pooling comprises 1800~frames (one minute at 30 fps). Furthermore, besides the average quality provided within the viewport throughout the considered temporal window ($q_{\text{window}}$), we assess the quality of the session computing the percentage of frames with a quality value higher than 80\% of the maximum quality ($f_{\text{window}}$ imposing $T_{\text{Q}}=0.8$).

\begin{figure}[t]
\centering
\subfloat[500~ms]{\includegraphics[width=0.9\columnwidth]{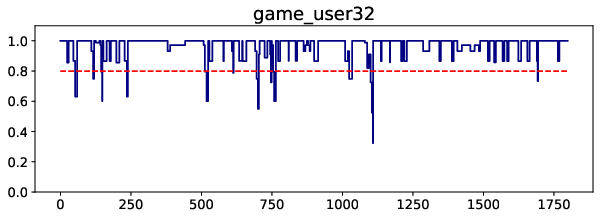}}\\
\subfloat[2000~ms]{\includegraphics[width=0.9\columnwidth]{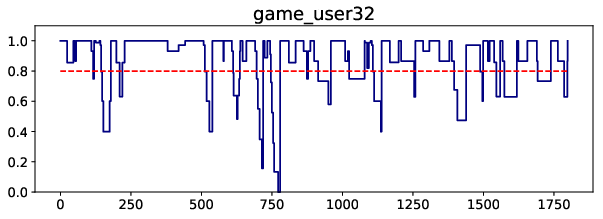}}\\
\subfloat[6000~ms]{\includegraphics[width=0.9\columnwidth]{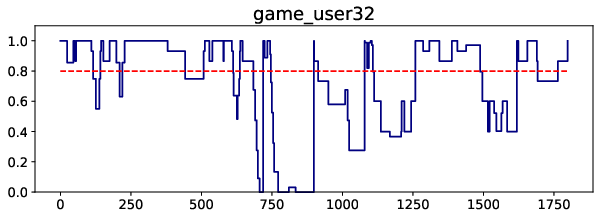}}
\caption{\label{fig:difSegsizes}{Evolution of the spatial quality pooling over time for user 32 and content 'game' using three different segment lengths. The dashed red line shows the 80\% threshold ($T_{\text{Q}}=0.8$)}}
\end{figure}

\subsection{Impact of the segment length}
For the first set of experiments, the described methodology has been used to compute the observed quality over time during a specific session as a function of the segment length. To that end, we have simulated the use of segments of three different lengths: 500~ms, 2000~ms and 6000~ms. Figure~\ref{fig:difSegsizes} shows the quality presented to the same user (user 32 in the dataset) visualizing the same content (content 'game' in the dataset) if different segment lengths were used. Additionally, the results of the two temporal quality pooling approaches proposed in Section~\ref{sec:Methodology} are included in Table~\ref{tab:difSegSizesTable}.

\begin{table}[ht]
\caption{\label{tab:difSegSizesTable}{Aggregated spatial and temporal quality pooling for user 32, content 'game' and different segment lengths. $T_{\text{Q}}=0.8$}}
\centering
\begin{tabular}{c|c|c|}
Segment length & $q_{\text{window}}$ & $f_{\text{window}}$  \\
\hline
500~ms & 0.9650 & 95.89\% \\
2000~ms & 0.8679 & 74.11\% \\
6000~ms & 0.7232 & 53.72\%
\end{tabular}
\end{table}

As can be observed, the longer the segment length, the lower the average quality, and the lower the percentage of time above the threshold. This is due to the slower adaptation of the viewport-oriented content to the new user's PoG.

\begin{figure}[t]
\centering
\subfloat[User 28, content 'coaster' (driven) ]{\includegraphics[width=0.9\columnwidth]{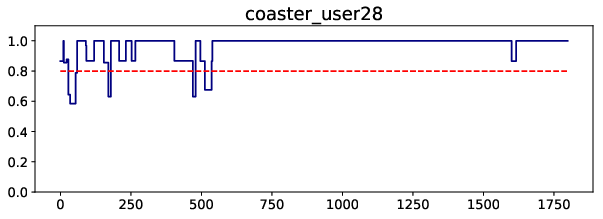}}\\
\subfloat[User 28, content 'game' (neither driven nor exploratory)]{\includegraphics[width=0.9\columnwidth]{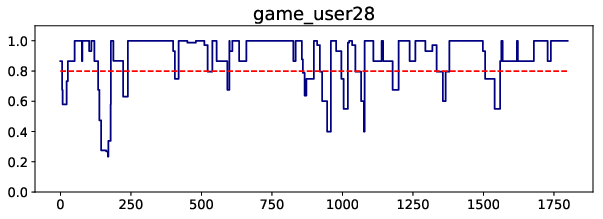}}\\
\subfloat[User 28, content 'landscape' (exploratory)]{\includegraphics[width=0.9\columnwidth]{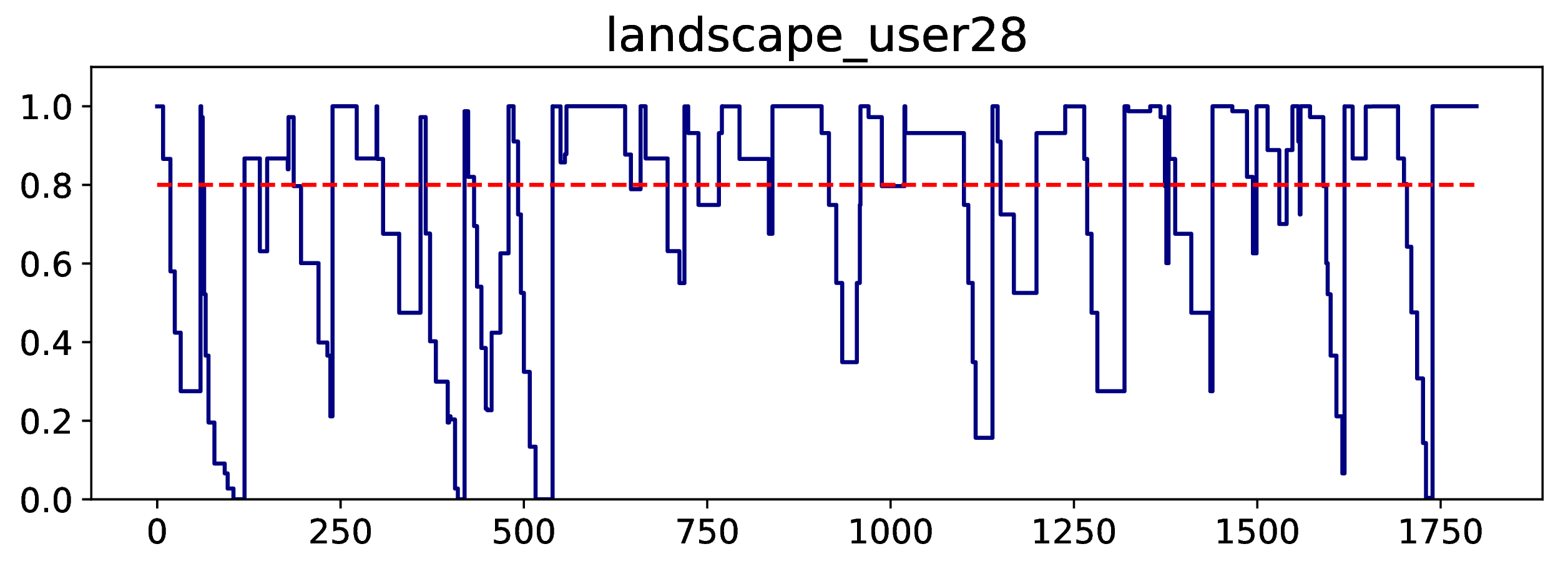}}
\caption{\label{fig:Evolution-difcontents}{Evolution of the spatial quality pooling over time for user 28, segment length 2000~ms, and three different contents. The dashed red line shows the 80\% threshold ($T_{\text{Q}}=0.8$)}}
\end{figure}

\subsection{Impact of the amount of exploration}
The amount of exploration, and thus of variation of the user PoG, during a session depends basically on two factors: the type of content (how driven or exploratory it is) and the user's own nature (how active or calm he/she is when viewing 360VR content). Therefore, the analysis of these elements is key to designers to properly build and tune encoding and transmission strategies to provide QoE.

\begin{table}[ht]
\caption{\label{tab:difContentsTable}{Aggregated spatial and temporal quality pooling for user 28, segment length of 2000~ms and different contents. $T_{\text{Q}}=0.8$}}
\centering
\begin{tabular}{c|c|c|}
Content & $q_{\text{window}}$ & $f_{\text{window}}$  \\
\hline
coaster & 0.9700 & 95.94\% \\
game & 0.8956 & 79.33\% \\
landscape & 0.8137 & 64.39\%
\end{tabular}
\end{table}

In this subsection, we use the proposed method to accurately study this dependency. First, we analyze the degree to which the nature of the content boosts exploration across the equirectangular image, which, as mentioned, can notably impact the quality perceived by users in viewport-adaptive schemes. To that end, we compute the spatial quality pooling along time obtained for a number sessions. Each session consists in the same user watching a different content. We have selected three sessions with rather different contents on the driven/exploratory dimension. The results for these sessions are included in Figure~\ref{fig:Evolution-difcontents}. Furthermore, we have included the aggregated spatial and temporal quality pooling to provide a global quality value per session. These results are included in Table~\ref{tab:difContentsTable}.

As expected, the more exploratory it is the content presented to the user, the more it encourages him/her to move, which causes more quality changes, as reflected in the figure. This moves result in a lower average quality, and a lower percentage of time above the threshold, as shown in the table.

\begin{figure}[t]
\centering
\subfloat[Curious user]{\includegraphics[width=0.9\columnwidth]{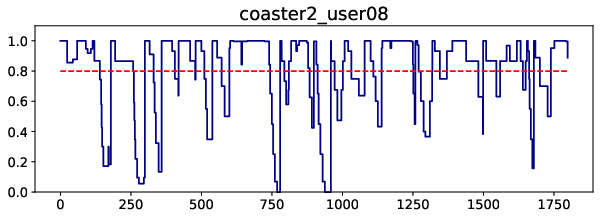}}\\
\subfloat[Medium user]{\includegraphics[width=0.9\columnwidth]{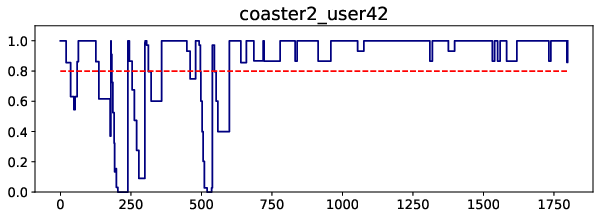}}\\
\subfloat[Quiet user]{\includegraphics[width=0.9\columnwidth]{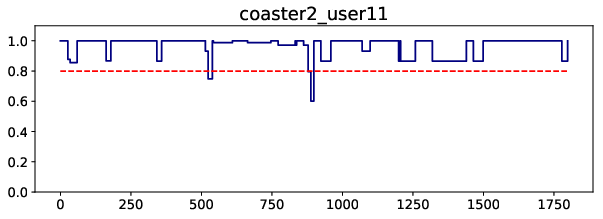}}
\caption{\label{fig:Evolution-difUsers}{Evolution of the spatial quality pooling over time for content 'coaster2', segment length 2000~ms, and three different users. The dashed red line shows the 80\% threshold ($T_{\text{Q}}=0.8$)}}
\end{figure}

\begin{table}[ht]
\caption{\label{tab:difUsersTable}{Aggregated spatial and temporal quality pooling for content 'coaster2', segment length of 2000~ms and different users. $T_{\text{Q}}=0.8$}}
\centering
\begin{tabular}{c|c|c|}
User & $q_{\text{window}}$ & $f_{\text{window}}$  \\
\hline
user 8 & 0.7999 & 68.22\% \\
user 42 & 0.8689 & 81.67\% \\
user 11 & 0.9654 & 98.00\%
\end{tabular}
\end{table}

Next, we study the influence of the user's behavior in the degree of exploration. So again we compute the spatial quality pooling along time obtained for a number of sessions. Each of these session consists in the content presented to a different user. Figure~\ref{fig:Evolution-difUsers} depicts the results for three users that could be classified as curious, medium and quiet. Furthermore, we have included the aggregated spatial and temporal quality pooling to provide a global quality value for each of these sessions. These results are included in Table~\ref{tab:difUsersTable}.

The results show that the more active the user is, the more he/she moves, and so the more quality changes along the session. As before, this type of session correlates with a lower average quality, and a lower percentage of time above the threshold.

\subsection{Summary of results}

\begin{table}[t!]
\renewcommand{\arraystretch}{1.3}
\caption{\label{tab:allQualityPooling}{Average temporal and spatial quality pooling per content and segment length. $T_{\text{Q}}=0.8$}}
\centering
\resizebox{\columnwidth}{!}{
\begin{tabular}{c|c|c|c|c|c|c}
\multirow{3}{*}{Content} & \multicolumn{6}{c}{Segment length}\\
\cline{2-7}
& \multicolumn{2}{c|}{500~ms} & \multicolumn{2}{c|}{2000~ms} & \multicolumn{2}{c}{6000~ms} \\
\cline{2-7}
& $q_{\text{window}}$      & $f_{\text{window}}$     & $q_{\text{window}}$      & $f_{\text{window}}$      & $q_{\text{window}}$      & $f_{\text{window}}$      \\ \hline
coaster                                                 & 0.9779           & 98.48\%           & 0.9250            & 89.81\%           & 0.8472            & 79.55\%           \\
coaster2                                                & 0.9787           & 98.47\%           & 0.9258            & 89.57\%           & 0.8396            & 76.85\%           \\
diving                                                  & 0.9753           & 98.53\%           & 0.9075            & 86.10\%           & 0.7619            & 63.45\%           \\
drive                                                   & 0.9697           & 97.50\%           & 0.8845            & 81.95\%           & 0.7609            & 64.43\%           \\
game                                                    & 0.9767           & 97.92\%           & 0.9041            & 85.84\%           & 0.8231            & 74.98\%           \\
landscape                                               & 0.9718           & 97.30\%           & 0.8707            & 78.21\%           & 0.7358            & 60.65\%           \\
pacman                                                  & 0.9787           & 97.80\%           & 0.9230            & 88.37\%           & 0.8440            & 77.61\%           \\
panel                                                   & 0.9670           & 97.33\%           & 0.8785            & 80.59\%           & 0.7014            & 57.58\%           \\
ride                                                    & 0.9751           & 97.87\%           & 0.9037            & 85.81\%           & 0.8075            & 70.37\%           \\
sport                                                   & 0.9731           & 98.28\%           & 0.9048            & 86.04\%           & 0.7787            & 64.49\%           \\ \hline
\multicolumn{1}{c|}{Average}                           & 0.9744  & 97.95\%  & 0.9028   & 85.23\%  & 0.7904  & 68.99\%  \\
\end{tabular}
}
\end{table}

Finally, we include the average aggregated temporal and spatial quality pooling per content and per segment length in Table~\ref{tab:allQualityPooling}. As expected, on average, there is a clear difference between the results obtained for different segment lengths, with shorter segments allowing a better adaptation and so a better global quality.

Focusing on the shortest segment length, 500~ms, we can observe that the values obtained after performing the temporal quality pooling do not vary much with the content. The reason is that the use of very short segments provides quick adaptation, thus preventing the user's PoG from moving much from its position at the beginning of the segment, when it was last updated, regardless of the nature of the content and, as a matter of fact, of the behavior of the user. However, the longer the segments used in the session, the lower the quality perceived by users on average and the higher the variations between contents. The overall quality drop is a consequence of the increasing time the user has to explore the scene before the viewport is updated, regardless of the content. The variability in the quality drop amplitude reflects the influence of the characteristics of the content in the amount of exploration. In this respect, the more exploratory the presented content, the lower the temporal quality pooling. Both tendencies reinforce the analysis presented in previous subsections.

\section{\label{sec:Conclusions}Conclusions}
The accurate assessment of the quality perceived by users throughout a 360VR video visualization session is key in the design of robust specific encoding and transmission strategies. In particular, the strict requirements to provide 360VR content with good quality have led to the development of many different viewport-adaptation strategies aiming at offering the best possible quality while saving bitrate. To properly evaluate these schemes, not only the saved bitrate, but also the quality of the portion of the scene actually presented through the HMD at all times should be consider. In this paper, we have proposed a methodology to accurately assess the quality inside the viewport around the user's point of gaze at every moment. This methodology has been made possible thanks to a complete analysis of the geometric relations involved in this particular environment, also detailed in the paper.

The proposed procedure is highly flexible and allows for any trade-off between accuracy and computational load. This is done by selecting the degree of approximations that best suits the specific requirements of the scenario. These options enable the use of the proposed methodology both offline and online, depending on the needs of the system.

Finally, we have shown its operation through a set of descriptive experiments. In particular, we have tested the effect of different essential factors on the observed quality, such as the length of the segments and the amount of movement of the user along the session. The analysis of the results validates the capability of the proposed methods to assess the quality perceived by users from different perspectives.

\bibliographystyle{IEEEtran}
\bibliography{bibliografia}



\end{document}